\documentclass[a4paper,11pt]{article}
\pdfoutput=1
\usepackage{jcappub}

\usepackage{amsmath}
\usepackage{bm}
\usepackage{xcolor,color,colortbl}
\usepackage[utf8]{inputenc}
\usepackage{hyperref}
\usepackage{url}
\usepackage{graphicx}
\usepackage{multirow,bigdelim}
\usepackage{amssymb}
\usepackage[amssymb]{SIunits}
\usepackage{multirow}
\usepackage{bbold}
\usepackage{verbatim}
\usepackage{comment}
\usepackage{caption}
\usepackage{physics}
\usepackage{subcaption}
\usepackage{orcidlink}

\usepackage{tikz}
\usetikzlibrary{positioning}

\definecolor{darkgreen}{HTML}{339933}

\usepackage{scalerel,tikz}
\usetikzlibrary{svg.path}
\definecolor{orcidlogocol}{HTML}{A6CE39}
\tikzset{orcidlogo/.pic={
 \fill[orcidlogocol] svg{M256,128c0,70.7-57.3,128-128,128C57.3,256,0,198.7,0,128C0,57.3,57.3,0,128,0C198.7,0,256,57.3,256,128z};
 \fill[white] svg{M86.3,186.2H70.9V79.1h15.4v48.4V186.2z}
 svg{M108.9,79.1h41.6c39.6,0,57,28.3,57,53.6c0,27.5-21.5,53.6-56.8,53.6h-41.8V79.1z M124.3,172.4h24.5c34.9,0,42.9-26.5,42.9-39.7c0-21.5-13.7-39.7-43.7-39.7h-23.7V172.4z}
 svg{M88.7,56.8c0,5.5-4.5,10.1-10.1,10.1c-5.6,0-10.1-4.6-10.1-10.1c0-5.6,4.5-10.1,10.1-10.1C84.2,46.7,88.7,51.3,88.7,56.8z};
}}
\newcommand\orcidicon[1]{\href{https://orcid.org/#1}{\mbox{\scalerel*{
\begin{tikzpicture}[yscale=-1,transform shape]
\pic{orcidlogo};
\end{tikzpicture}
}{|}}}}

\title{The Marked Power Spectrum as a Practical Bispectrum Measure for Galaxy Redshift Surveys}

\author[a,b,c]{Haruki Ebina~\orcidicon{0000-0002-1080-0955},}
\author[a,b,c]{Martin White~\orcidicon{0000-0001-9912-5070},}
\author[c]{and Edmond Chaussidon~\orcidicon{0000-0001-8996-4874}}
\affiliation[a]{Department of Physics, University of California, Berkeley, CA 94720, USA}
\affiliation[b]{Berkeley Center for Cosmological Physics, UC Berkeley, CA 94720, USA}
\affiliation[c]{Lawrence Berkeley National Laboratory, One Cyclotron Road, Berkeley, CA 94720, USA}

\emailAdd{ebina@berkeley.edu}

\abstract{
Modern datasets have the precision necessary to uncover new information by including higher-order, non-Gaussian information into cosmological inference. The marked power spectrum offers access to such information while preserving the structure of two-point correlators. This approach to higher-order statistics has the advantage that many modeling questions can directly benefit from progress already made in standard cosmological analyses using the power spectrum and correlation function, while increasing the data vector size negligibly and retaining much of the degeneracy-breaking power of the bispectrum through a particular data compression. In this work, we first restructure the marked power spectrum to isolate its higher-order information and demonstrate its ability to break parameter degeneracies. We then investigate the effect of survey geometry on the marked power spectrum and find that a treatment similar to that of the power spectrum is sufficient. Additionally, we investigate the perturbative modeling and covariance structure of the marked power spectrum, shedding light on its degeneracy breaking power and cross-covariance with the power spectrum. Finally, we demonstrate that the cosmology dependence of the marked power spectrum is smooth, indicating that cosmological inference is possible by modeling the cosmology dependence through interpolation rather than analytical modeling. 
}

\begin{document}
\maketitle
\flushbottom

\section{Introduction}
\label{sec:introduction}

Large-scale structure offers a unique probe into the evolution of the universe by tracing the gravitational growth of the density field, enabling local observers to learn about cosmology, fundamental physics, and structure formation \cite{Bernardeau02,Baumann22,Ferraro22}. 
For redshift surveys to date, including the ongoing Dark Energy Spectroscopic Instrument (DESI) \cite{DESI-DR1}, the majority of the information has come from the two-point function, either in Fourier space (the power spectrum) or configuration space (the correlation function) \cite{DESI-DR2,DESI24-V}. The near-Gaussianity of the density field on large scales makes this an efficient compression of information. However, this compression is not entirely lossless and with rapidly increasing observational precision, attention has turned to extracting the small amount of additional information present in higher-order statistics on quasi-linear scales.
There are many means to access this additional information.  Higher-order $n$-point functions, such as the bispectrum \cite{Peebles75,Fry82} and trispectrum \cite{Fry78,Hu01}, are the most straightforward generalizations of the two-point function, while a variety of alternative statistics exist, including skew spectra \cite{Schmittfull15,Hou24}, wavelet scattering transforms \cite{Cheng20,Valogiannis22,Eickenberg22,Cheng24}, density-split statistics \cite{Gruen16,Friedrich18,Paillas21,Paillas24} and the filter-squared bispectrum \cite{Harscouet24,Verdiani25}.
Many of these methods have been compared in the recent ``Beyond 2-pt Challenge'' \cite{Krause24}.

While straightforward in principle, higher-order $n$-point functions pose several practical challenges: computational complexity, large data vectors, requirements on covariances, window functions, sensitivity to systematics or artifacts from the survey operations \cite{Takahashi20,Philcox22c,Giri22,Giri23}. Since much of the new information provided by higher order statistics comes from their ability to break degeneracies between ``nuisance parameters'' \cite{Bernardeau97,Pires12,Hahn21}, that are themselves degenerate with cosmological parameters, there is reason to believe that alternatives that are practically easier to handle may achieve similar performance.

In particular, marked spectra can offer a practical means of breaking degeneracies among parameters in our theoretical models while being theoretically well controlled, having small data sizes and making the most use of the existing survey infrastructure \cite{White16}. The perturbative modeling of the marked power spectrum is on the same footing as the existing power spectrum models, with the marked spectrum having a straightforward relation to the bispectrum as well. Extensive infrastructure exists to handle survey practicalities and systematics in two-point functions and marked spectra can make use of much of this. Density field estimation, an essential component of marked spectra observation, is also well developed, since it is used as a key part of analyses of baryon acoustic oscillations. Further, marks that are low-order in the smoothed overdensity field, with smoothing radii that are larger than the non-linear scale, are especially amenable to an analytic treatment. For all of these reasons, properly constructed marked power spectra have been suggested as a means of enhancing cosmological inference \cite{Ebina24,Cowell25}.

Previously, marked spectra have been proposed as a means of testing cosmology by up-weighting underdense regions with a particular inverse-weighting mark, which enhances sensitivity to modified gravity effects or massive neutrinos \cite{White16,Massara21,Massara23}. More recently, perturbative models for marked spectra have been developed \cite{Philcox20,Philcox21,Ebina24} and the possibility of using different forms of the mark has been investigated \cite{Ebina24,Cowell24}. Ref.~\cite{Ebina24} in particular found that modeling is easier for marks that were low-order polynomials in the smoothed overdensity and have demonstrated that marked spectra is capable of improving cosmological inference by breaking degeneracies among parameters in the theoretical models. Following these works, we develop a method to incorporate survey window functions into the marked power spectrum, which is essential for applying marked spectra to real survey data. Furthermore, we restructure the marked power spectrum to isolate the higher-order information and demonstrate its ability to break parameter degeneracies.

This paper is organized as follows.  We begin in \S\ref{sec:mps} with a review of the marked power spectrum and its perturbative modeling, and restructure the spectrum to isolate the higher-order information. In \S\ref{sec:covariance} we describe the covariance structure of marked power spectra and quantify the amount of new information as compared to the power spectrum. In \S\ref{sec:window} we describe how to measure the marked spectrum on survey data and incorporate survey window functions in a similar manner to the power spectrum. In \S\ref{sec:periodicbox} and \S\ref{sec:cutsky} we validate the modeling on periodic box and cutsky simulations, respectively. In \S\ref{sec:AP} we show how to accurately and cheaply incorporate the fiducial cosmology dependence in cosmology inference. Finally, we present our conclusions in \S\ref{sec:conclusion}.

\section{Marked Power Spectrum}
\label{sec:mps}

The marked power spectrum (MPS) offers a principled, controlled method to include higher-order information beyond the power spectrum. MPS was initially introduced in cosmology\footnote{Preceding work to this was mostly in the context of astronomy, where marks were defined with observables, such as luminosity, rather than density \cite{Beisbart00,Sheth05}.} to look for modified gravity signals by up-weighting void (underdense) regions \cite{White16}. This particular mark (weighting) has been adopted by subsequent studies that predicted significant improvement in cosmological constraints, in particular neutrino mass \cite{Massara21,Massara23}. Recent work \cite{Philcox20,Philcox21,Ebina24} has shown that a perturbative modeling of the MPS is possible without introducing UV uncertainties beyond those already in the power spectrum. Refs.~\cite{Ebina24,Cowell24} investigate the possibility of using different forms of the mark beyond that was initially proposed, where ref.~\cite{Ebina24} in particular found that modeling is easier for marks that were low-order polynomials in the smoothed overdensity. In the following text we introduce the theoretical set up of the MPS, focusing on the linear mark investigated in ref.~\cite{Ebina24}. While the inverse-weighting mark \cite{White16} can be described in an identical, perturbative formalism to what we employ in this work \cite{Philcox21}, the infinite Taylor expansion of the mark creates unpredictable higher-loop contributions, making the pertubative modeling practically challenging. To one-loop power spectrum order ($\mathcal{O}(\delta^4)$), we believe that the linear formalism captures all new information introduced by any mark, as we will describe below. 

As we will frequently transition between configuration and Fourier space, it will be useful to define the following notation
\begin{equation}
    \expval{f_1(\mathbf{x}_1)f_2(\mathbf{x}_2)}_{\rm FT}(\mathbf{k}) = \int d\mathbf{x}\ e^{-i\mathbf{k}\cdot\mathbf{x}}\left.\expval{f_1(\mathbf{x}_1)f_2(\mathbf{x}_2)}\right|_{\mathbf{x}=\mathbf{x}_1-\mathbf{x}_2}
\end{equation}
where $f_1$ and $f_2$ are fields defined in configuration-space and we assume that the two-point correlator depends only on the separation vector between the two fields $\mathbf{x}=\mathbf{x}_1-\mathbf{x}_2$. To simplify later expressions we shall also adopt the notation
\begin{equation}
    \int_{\mathbf{k}_1\cdots\mathbf{k}_n}
    = \int\prod_{i=1}^n \frac{d^3k_i}{(2\pi)^3}
    \quad .
\end{equation}

\subsection{Definition and modeling}

The marked density field is defined by weighting the galaxy density field $\rho_g$ by a mark field, $m$,
\begin{equation}
    \rho_M(\mathbf{x}) = m\left[\delta_{g,R}(\mathbf{x})\right] \rho_g(\mathbf{x})
\end{equation}
taken to be a functional of the smoothed galaxy overdensity field, $\delta_{g,R}$.  We define $\delta_{g,R}(k)=W_R(k)\delta_g(k)$ with a Gaussian kernel $W_R(k)=\exp{-k^2 R^2/2}$\footnote{The possibility of using a non-Gaussian kernel has been explored in past work \cite{Massara21,Philcox20} and more recently in ref.~\cite{Gao25}, but will be outside the scope of this study.}. 
Expanding the mark field in powers of $\delta_{g,R}$
\begin{equation}
     m(\bm{x}) = \sum_n \widetilde{C}_n \delta_{g,R}^n(\bm{x})
\end{equation}
one finds that only terms with $n\le3$ appear when perturbatively modeling the power spectrum to one-loop order \cite{Ebina24,Philcox21}. For a controlled modeling of correlators of $\rho_M$, it is useful to consider marks that are low order in $\delta_{g,R}$ \cite{Ebina24}. In this work, we follow ref.~\cite{Ebina24} and focus on marks that are linear in $\delta_{g,R}$
\begin{equation}
     \frac{m(\bm{x})}{\bar{m}} = C_0 + C_1\delta_{g,R}(\bm{x})
\end{equation}
making the marked overdensity field 
\begin{align}
    \delta_M &= (m/\bar{m})\left(\delta_g + 1\right) - 1 \\
    &= (C_0 + C_1\delta_{g,R})(\delta_g+1)-1
\end{align}
where $\bar{m}=\expval{\rho_M}/\expval{\rho_g}$ is an overall normalization that can be measured from simulations or data and we have defined $\widetilde{C}_n=\bar{m}\,C_n$.
We shall be exclusively concerned with the Fourier-space two-point function of $\delta_M$, which is known as the marked power spectrum (MPS).  Being a two-point function, it can be efficiently computed and manipulated using the same infrastructure as is commonly used to measure the power spectrum in galaxy surveys.  A key quality of the MPS is that it introduces no additional UV-divergences than the power spectrum, as the smoothing on all `additional' fields prevents UV-divergent contact terms (zero-lag contractions, e.g.~$\delta_g^2(\mathbf{x})\supset\int P(p) dp$).  This places the MPS on an equal footing with the power spectrum from the perspective of modeling. 

Previous studies on the MPS \cite{Ebina24,Philcox21, White16} have focused directly on the two-point correlator of the marked overdensity field $\mathcal{M} = \expval{\delta_M^2}$ which, due to the constant term in each $\delta_g$, inevitably includes two-point information already available from the power spectrum \cite{Ebina24,Philcox21}. Namely for two fields with marks $m^{a}$ and $m^{b}$,
\begin{align}
    \mathcal{M} &= \expval{ \delta_M^{a} \delta_M^{b}}
    := \expval{C_0^a C_0^b \delta_g^2 + (C_0^a C_1^b + C_0^b C_1^a ) \delta_g \delta_{g,R} + C_1^a C_1^b\delta_{g,R}^2} + M
    \label{defMorig}
\end{align}
where the first set of terms is merely a collection of power spectrum information, with $\expval{\delta_g^2}=P(k,\mu)$, $\expval{\delta_g \delta_{g,R}}=W_R(k) P(k,\mu)$ and $\expval{\delta_{g,R}^2}=W_R^2(k)P(k,\mu)$, and we have (re-)defined $M$ to be the collection of terms involving higher-point information. From here on, we will refer to $M$ as the \textit{marked power spectrum}\footnote{One can equally use a convention where one weighs the overdensity $\delta_g$ instead of $\rho_g$, making the marked overdensity $\bar{m}\delta_M=m\delta_g$, as in ref.~\cite{Cowell24}. Both conventions produce the spectra studied in this work, but are not in general equal.}. We refer the reader to Appendix \ref{sec:b2pinfo} for discussion on the two-point and beyond two-point information content in $\mathcal{M}$.

For any mark, the MPS $M$ can be decomposed into three- and four-point correlators \cite{Philcox21,Ebina24}. 
For a linear mark, this decomposition is
\begin{align}
    M &= \left[C_0^a C_1^b + C_0^b C_1^a\right]
         \expval{ \delta_{g,R}(\mathbf{x}_1)\,\delta_g(\mathbf{x}_1)\,\delta_g(\mathbf{x}_2) }
         \nonumber \\
      &\quad + 2 C_1^a C_1^b\,
         \expval{ \delta_{g,R}(\mathbf{x}_1)\,\delta_g(\mathbf{x}_1)\,\delta_{g,R}(\mathbf{x}_2) } 
         + C_1^a C_1^b\,
         \expval{ \delta_{g,R}(\mathbf{x}_1)\,\delta_g(\mathbf{x}_1)\,\delta_{g,R}(\mathbf{x}_2)\,\delta_g(\mathbf{x}_2) }
    \label{eqn:defM}
\end{align}
where the freedom in the coefficients shows our ability to access different terms by choice of mark(s). To access the first three-point correlator, we can choose one mark to be a constant, e.g.~a cross-spectrum between the unmarked field ($m=1$, so $\rho_M=\rho_g$) and $m=1+\delta_{g,R}$. In fact, any cross-spectrum between the unmarked field and a linear mark isolates the first term. 
To access the second and third terms, one must choose the spectrum between two linear marks, e.g.~the auto-spectrum of $m=1+\delta_{g,R}$, but to isolate these terms, one will need to null the first term, by e.g.~the cross-spectrum between $1+\delta_{g,R}$ and $1-\delta_{g,R}$.
To the extent that, at one-loop power spectra order ($\mathcal{O}(\delta^4)$), the four-point correlators merely contribute the disconnected trispectrum (convolutions of the power spectrum), the three-point correlation terms of MPS contain all of the new (beyond two point) information introduced by the MPS. 

Thus, in this work, we mainly focus on the cross-spectrum between the unmarked field $\delta_g$ and the mark $m=1+\delta_{g,R}$ (further discussion of alternative marks can be found in Appendix \S\ref{sec:stability}).  This leaves us with
\begin{equation}
    M = \frac{1}{\bar{m} } \expval{ \delta_{g,R}(\mathbf{x}_1)\,\delta_g(\mathbf{x}_1)\,\delta_g(\mathbf{x}_2) }
\end{equation}
in configuration space.  In Fourier space
\begin{equation}
    M(\mathbf{k}) = \frac{1}{\bar{m} } \langle [\delta_{g,R}\delta_g](\mathbf{k})\delta_g(-\mathbf{k})\rangle
    = \frac{1}{\bar{m} } \int \frac{d^3p}{(2\pi)^3} W_R(p)B(\mathbf{p},\mathbf{k}-\mathbf{p},-\mathbf{k})
\label{eqn:bispectrum-integral}
\end{equation}
where $B$ is the galaxy bispectrum. The MPS is an integral of the bispectrum with the integration limited to $p\sim R^{-1}$ by (one power of) the smoothing kernel. 

This form of the MPS makes apparent the similarity of MPS with the skew spectrum, a wide class of composite-field two-point correlators \cite{Cooray01,Munshi10,Schmittfull15,Moradinezhad20}. 
One of the first and simplest skew spectra, $\expval{[\delta_{g,R}^2](\mathbf{k})\delta_g(-\mathbf{k})}$, merely differs from the MPS expression above by an extra smoothed field in the convolution. This indicates that the information obtained between the MPS and skew spectra are similar, but likely greater from the MPS due to the lack of smoothing. On the other hand, the extra smoothing of the skew spectra likely aids the modeling and control of non-linearities. 
There are also practical differences in the measurement and window treatment.
We will elaborate on this further in \S\ref{sec:window}. In general, an investigation into the differences between various composite-field two-point correlators would be interesting, but is beyond the scope of this work. 

For the remainder of this work, we will not explicitly write the factor of $\bar{m}$ as it is only an overall factor of $\approx 1$. This corresponds to setting the mark to $m=1+\bar{m} \delta_{g,R}$ and is well-defined, however one cannot know $\bar{m}$ without selecting a dataset and hence this mark would not be fit for a general theoretical discussion. 

\subsection{Perturbative modeling}
\label{sec:PTintro}

Similarly to the power spectrum, the MPS can be modeled perturbatively in powers of the linear overdensity field $\delta_L$ \cite{Ebina24,Philcox21}. In Eulerian Perturbation Theory (EPT; \cite{Bernardeau02,Ivanov22b}), the equations of motion are constructed based on the assumption that cold dark matter and baryons behave as perfect, pressure-less fluids. These equations of motions can then be solved perturbatively, predicting the nonlinear matter overdensity field $\delta=\rho/\bar{\rho}-1$ and velocity divergence $\theta=\nabla\cdot\mathbf{v}$ order-by-order.  In the Einstein-de Sitter limit, writing $\delta=\sum_{n=1}^\infty D^n\delta^{(n)}$ and $\theta=\sum_{n=1}^\infty D^n\theta^{(n)}$, with $D$ the linear growth factor, one finds \cite{Bernardeau02,Ivanov22b}
\begin{equation}
    \delta^{(n)}(\mathbf{k}) = \int_{\mathbf{k}_1\dots\mathbf{k}_n} (2\pi)^3 \delta_D\left(\sum_i \mathbf{k}_i -\mathbf{k}\right) \delta_L(\mathbf{k}_1)\cdots\delta(\mathbf{k}_n) F_n(\mathbf{k}_1,\dots,\mathbf{k}_n,\mathbf{k})
\end{equation}
where $F_n$ is a function of the wavevectors that can be computed via recurrence, and for which we give explicit expressions for $n=1$ and 2 below.  A similar expression holds for $\theta^{(n)}$ with the substitution of $G_n$ for $F_n$. 
These predictions for the matter field are then translated into observable, redshift-space galaxy field by applying redshift-space distortions and using the bias expansion
\begin{equation}
    \delta_g = b_1 \delta_m + b_2 (\delta_m^2-\expval{\delta_m^2}) + b_s (s^2 - \expval{s^2}) + b_3 \mathcal{O}_3 + \cdots
\end{equation}
where we expand in the underlying matter and shear field, and the expansion terms are limited by symmetry and perturbative order. The final perturbative solution to the redshift-space galaxy overdensity field then becomes \cite{Bernardeau02,Ivanov22b} 
\begin{equation}
    \delta_g^{(n)}(\mathbf{k}) = \int_{\mathbf{k}_1\dots\mathbf{k}_n} (2\pi)^3 \delta_D\left(\sum_i \mathbf{k}_i -\mathbf{k}\right) \delta_L(\mathbf{k}_1)\cdots\delta(\mathbf{k}_n) Z_n(\mathbf{k}_1,\dots,\mathbf{k}_n,\mathbf{k})
    \label{eqn:deltagn}
\end{equation}
where the kernels $Z_n$ are defined by a combination of density kernels, $F_n$, velocity kernels, $G_n$, and bias terms. The first few kernels relevant for this work are \cite{Bernardeau02}
\begin{align}
    Z_1(\mathbf{k}_1) &= b_1 + f\mu_1^2 \\
    Z_2(\mathbf{k}_1,\mathbf{k}_2) &= b_1 F_2(\mathbf{k}_1,\mathbf{k}_2) + f\mu_k^2G_2(\mathbf{k}_1,\mathbf{k}_2) + \frac{b_2}{2} + b_s\left(\frac{(\mathbf{k}_1\cdot\mathbf{k}_2)^2}{k_1^2k_2^2}-\frac{1}{3} \right) \nonumber \\ 
    &+ \frac{fk\mu_k}{2}\left[\frac{\mu_1}{k_1}(b_1+f\mu_2^2) + \frac{\mu_2}{k_2}(b_1 + f\mu_1^2) \right]
\end{align}
with $\mathbf{k}=\mathbf{k}_1+\mathbf{k}_2$, $\mu_k=\widehat{k}\cdot\widehat{n}$,  $\mu_n=\widehat{k}_n\cdot\widehat{n}$, and the second-order density and velocity kernels are
\begin{align}
    F_2(\mathbf{k}_1, \mathbf{k}_2) &= \frac{5}{7} + \frac{2}{7}\left(\widehat{k}_1\cdot\widehat{k}_2\right)^2 + \frac{1}{2}\left(\widehat{k}_1\cdot\widehat{k}_2\right) \left(\frac{k_1}{k_2}+ \frac{k_2}{k_1}\right)     
    \label{eqn:F2} \\ 
    G_2(\mathbf{k}_1, \mathbf{k}_2) &= \frac{3}{7} + \frac{4}{7}\left(\widehat{k}_1\cdot\widehat{k}_2\right)^2 + \frac{1}{2}\left(\widehat{k}_1\cdot\widehat{k}_2\right) \left(\frac{k_1}{k_2}+ \frac{k_2}{k_1}\right) .
    \label{eqn:G2}
\end{align}
It is then clear that the contributions to the (unmarked) power spectrum up to one-loop ($\mathcal{O}(\delta_L^4)$) order only arise from three distinct contributions
\begin{equation}
    P(\mathbf{k}) = P_{11}(\mathbf{k}) + 2 P_{13}(\mathbf{k}) + P_{22}(\mathbf{k})
\end{equation}
with
\begin{equation}
    P_{11}=\expval{\left(\delta_g^{(1)}\right)^2}, \qquad 
    P_{13}=\expval{\delta_g^{(1)}\delta_g^{(3)}}, \qquad
    P_{22}=\expval{\left(\delta_g^{(2)}\right)^2}
\end{equation}
where the first term is the tree-level power spectrum \cite{Kai87} and the latter two are one-loop contributions \cite{Chen20a}. Note that the modeling here is limited to large scales by the validity of the original equations of motion set up for $\delta$ and $\theta$. To go beyond this we take an effective field theory (EFT) approach, where one integrates over the small-scales that are not described by the large-scale theory. We will discuss this, along with the treatment of field stochasticity later in \S\ref{sec:nuisance}. 
We refer the reader to ref.~\cite{Bernardeau02} for a detailed discussion of perturbative modeling and ref.~\cite{Ivanov22b} for a recent review of EFT in this context.

\begin{figure}
\centering
\begin{tikzpicture}[
deltagR/.style={circle, draw=red!60, fill=red!20, very thick, minimum size=10mm},
deltag/.style={rectangle, draw=red!60, fill=red!20, very thick, minimum size=10mm},
]
\node[deltag]      (maintopic)  at (0,0) {$\delta_g^{(2)}$};
\node[deltagR]        (uppercircle) at (0,2) {$\delta_{g,R}^{(1)}$};
\node[deltag]      (rightsquare)  at (2,0) {$\delta_g^{(1)}$};
\node[] (blank1) at (1,3.) {};
\node[] (blank2) at (1,-1) {};
\node[] (label1) at (0,3.) {$\delta_M^{(3)}(\mathbf{x}_1)$};
\node[] (label2) at (2,3.) {$\delta_g^{(1)}(\mathbf{x}_2)$};

\draw[very thick] (uppercircle.south) -- (maintopic.north);
\draw[very thick] (maintopic.east) -- (rightsquare.west);
\draw[dashed] (blank1.south) -- (blank2.north);
\end{tikzpicture}
\hspace{5em}
\begin{tikzpicture}[
deltagR/.style={circle, draw=red!60, fill=red!20, very thick, minimum size=10mm},
deltag/.style={rectangle, draw=red!60, fill=red!20, very thick, minimum size=10mm},
]
\node[deltagR]      (maintopic)  at (0,0) {$\delta_{g,R}^{(2)}$};
\node[deltag]        (uppercircle) at (0,2) {$\delta_{g}^{(1)}$};
\node[deltag]      (rightsquare)  at (2,0) {$\delta_g^{(1)}$};
\node[] (blank1) at (1,3.) {};
\node[] (blank2) at (1,-1) {};
\node[] (label1) at (0,3.) {$\delta_M^{(3)}(\mathbf{x}_1)$};
\node[] (label2) at (2,3.) {$\delta_g^{(1)}(\mathbf{x}_2)$};

\draw[very thick] (uppercircle.south) -- (maintopic.north);
\draw[very thick] (maintopic.east) -- (rightsquare.west);
\draw[dashed] (blank1.south) -- (blank2.north);
\end{tikzpicture}
\hspace{5em}
\begin{tikzpicture}[
deltagR/.style={circle, draw=red!60, fill=red!20, very thick, minimum size=10mm},
deltag/.style={rectangle, draw=red!60, fill=red!20, very thick, minimum size=10mm},
]
\node[deltag]      (maintopic)  at (0,0) {$\delta_g^{(1)}$};
\node[deltagR]        (uppercircle) at (0,2) {$\delta_{g,R}^{(1)}$};
\node[deltag]      (rightsquare)  at (2,1) {$\delta_g^{(2)}$};
\node[] (blank1) at (1,3.) {};
\node[] (blank2) at (1,-1) {};
\node[] (label1) at (0,3.) {$\delta_M^{(2)}(\mathbf{x}_1)$};
\node[] (label2) at (2,3.) {$\delta_g^{(2)}(\mathbf{x}_2)$};

\draw[very thick] (uppercircle.east) -- (rightsquare.west);
\draw[very thick] (maintopic.east) -- (rightsquare.west);
\draw[dashed] (blank1.south) -- (blank2.north);
\end{tikzpicture}
\caption{The three diagrams contributing to higher-order information in the MPS for a cross-spectrum between marked field $\delta_M$ at $\mathbf{x}_1$ and unmarked field $\delta_g$ at $\mathbf{x}_2$. The red squares and circles represent the raw and smoothed overdensity operators $\delta_g$ and $\delta_{g,R}$, respectively. The operators on the left of each diagram belong to the marked field, $\delta_M$, and those on the right belong to the unmarked field, $\delta_g$.  The left two diagrams contribute to $M_{13}$, whereas the rightmost diagram contributes to $M_{22}$. Notice that all diagrams include a $\delta^{(2)}\supset Z_2$, giving rise to a ``leading-order'' dependence on $b_2$ and $b_s$ that will be important later.}
\label{fig:PTdiagram}
\end{figure}
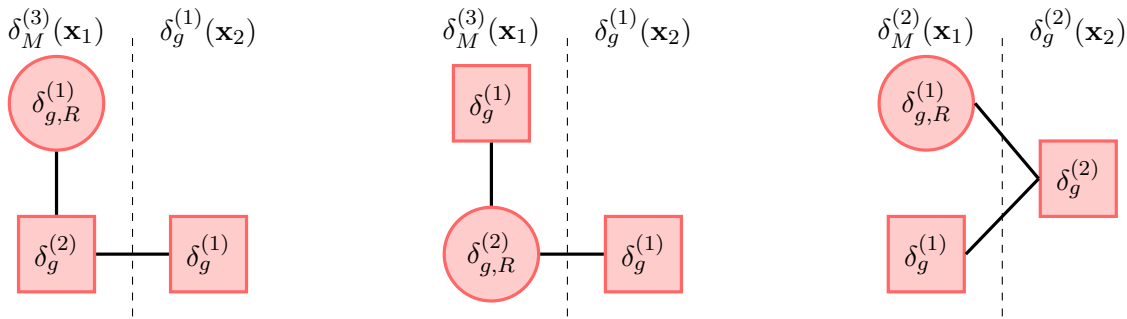

To extend the perturbative modeling to the MPS, it is useful to note that the smoothing of the overdensity field $\delta_{g,R}(\mathbf{k}) = W_R(k) \delta_g(\mathbf{k})$ simply extends to a smoothing on the perturbative solutions, i.e.~$\delta_{g,R}(\mathbf{k}) = \sum_{n=1}^\infty W_R(k)\delta_g^{(n)}(\mathbf{k})$. This alone is sufficient to model the MPS, modulo stochastic and small-scale dynamics (EFT) terms that we describe in \S\ref{sec:nuisance}. Recalling the expression for $M$ (Eqn.~\ref{eqn:defM}) the lowest order (one-loop in power spectrum, $\mathcal{O}(\delta_L^4)$) contributions then become
\begin{align}
\expval{ \delta_{g,R,1}\,\delta_{g,1}\,\delta_{g,2} }
    &= \expval{ \delta_{g,R,1}^{(1)}\,\delta_{g,1}^{(2)}\,\delta_{g,2}^{(1)} }
        + \expval{ \delta_{g,R,1}^{(2)}\,\delta_{g,1}^{(1)}\,\delta_{g,2}^{(1)} }
        + \expval{ \delta_{g,R,1}^{(1)}\,\delta_{g,1}^{(1)}\,\delta_{g,2}^{(2)} }, \\
\expval{ \delta_{g,R,1}\,\delta_{g,1}\,\delta_{g,R,2} }
    &= \expval{ \delta_{g,R,1}^{(1)}\,\delta_{g,1}^{(2)}\,\delta_{g,R,2}^{(1)} }
        + \expval{ \delta_{g,R,1}^{(2)}\,\delta_{g,1}^{(1)}\,\delta_{g,R,2}^{(1)} }
        + \expval{ \delta_{g,R,1}^{(1)}\,\delta_{g,1}^{(1)}\,\delta_{g,R,2}^{(2)} }, \\
\expval{ \delta_{g,R,1}\,\delta_{g,1}\,\delta_{g,R,2}\,\delta_{g,2} }
    &= \expval{ \delta_{g,R,1}^{(1)}\,\delta_{g,1}^{(1)}\,\delta_{g,R,2}^{(1)}\,\delta_{g,2}^{(1)} }
\end{align}
where the subscripts on the fields denote their spatial coordinates. The terms with three fields are all different compressions of the tree-level bispectrum $B_{112}= \expval{\delta_g^{(1)}\delta_g^{(1)}\delta_g^{(2)}}$ and the terms with four fields are convolutions of the tree-level power spectrum $P_{11}$. These terms can also be reorganized by the order of $\delta_M^{(n)}$ they belong to, i.e.
\begin{align}
    M &= 2 M_{13} + M_{22} \\
    2M_{13} &= \left[C_0^a C_1^b + C_0^b C_1^a\right] 
    \left( \expval{ \delta_{g,R,1}^{(1)}\,\delta_{g,1}^{(2)}\,\delta_{g,2}^{(1)} }
    + \expval{ \delta_{g,R,1}^{(2)}\,\delta_{g,1}^{(1)}\,\delta_{g,2}^{(1)} }  \right) \nonumber \\ 
    &\quad 
    + 2 C_1^a C_1^b \left(\expval{ \delta_{g,R,1}^{(1)}\,\delta_{g,1}^{(2)}\,\delta_{g,R,2}^{(1)} }
    + \expval{ \delta_{g,R,1}^{(2)}\,\delta_{g,1}^{(1)}\,\delta_{g,R,2}^{(1)} }\right) \\
    M_{22} &= \left[C_0^a C_1^b + C_0^b C_1^a\right] \expval{ \delta_{g,R,1}^{(1)}\,\delta_{g,1}^{(1)}\,\delta_{g,2}^{(2)} } \nonumber \\
    &\quad + C_1^a C_1^b \left(2 \expval{ \delta_{g,R,1}^{(1)}\,\delta_{g,1}^{(1)}\,\delta_{g,R,2}^{(2)} } 
    + \expval{ \delta_{g,R,1}^{(1)}\,\delta_{g,1}^{(1)}\,\delta_{g,R,2}^{(1)}\,\delta_{g,2}^{(1)} }\right) .
\end{align}
Note again that for this work, we will isolate the contribution $\left[C_0^a C_1^b + C_0^b C_1^a\right]\expval{ \delta_{g,R,1}\delta_{g,1}\delta_{g,2} }$ by choosing to observe the cross-spectrum between the unmarked field and marked field with mark $m=1+\delta_{g,R}$, as it contains the least amount of smoothing while accessing information in the bispectrum directly. The contractions that contribute to this term can be diagrammatically described as shown in Fig.~\ref{fig:PTdiagram}.
We refer the reader to refs.~\cite{Ebina24,Philcox21} for detailed discussions of the subtleties in the perturbative modeling of the MPS ($\mathcal{M}$). 

\subsection{Nuisance Parameters}
\label{sec:nuisance}

The equations of motion central to the perturbative modeling above hold only for large-scale dynamics where the ideal assumptions about matter are approximately correct. At small-scales, gravitational collapse and structure formation induce effects beyond the model. These effects are incorporated into the large-scale theory by the effective field theory (EFT) approach, where one integrates over the the small-scales and introduce the maximal degrees of freedom allowed by symmetry. 
In addition, one needs to account for the stochasticity $\epsilon$ of the density field. These effects together introduce the counterterm corrections 
\begin{equation}
    P(k,\mu) \supset \sum_{n=0}^{2} \alpha_{2n} \mu^{2n} k^2 P_L(k)
\end{equation}
and stochastic terms
\begin{equation}
    P(k,\mu) \supset \sum_{n=0}^{2} N_{2n} (k\mu)^{2n}
\end{equation}
where $N_0=N$ is the shotnoise and $N_2$, $N_4$ include the fingers-of-god \cite{Chen20a}. 

While the MPS can largely be described using nuisance parameters already present in the modeling of the power spectrum, the new density contractions in the MPS implies that there is a set of new nuisance parameters introduced \cite{Ebina24}. The only parameters that are included at one-loop power spectrum order (tree-level bispectrum order; $\mathcal{O}(\delta_L^4)$) are the stochastic terms $B_\mathrm{shot}$ and $A_\mathrm{shot}$, which stem from the tree-level bispectrum stochastic terms \cite{Ivanov22a,Bakx25,Chudaykin25a}
\begin{equation}
    B \supset \left[\left(\frac{B_{\rm shot}}{\bar{n}} b_1 + Nf\mu^2\right)Z_1(\bm{k_1}) P_L(k_1) + {\rm cycl.} \right] + \frac{A_\mathrm{shot}}{\bar{n}^2}
    \label{eqn:stochastic}
\end{equation}
The contributions proportional to $B_{\rm shot}$, $N$, and $A_\mathrm{shot}$ arise from the products $\expval{ (\delta) (\epsilon)(\epsilon \delta)}=\expval{\epsilon^2}\expval{\delta^2}$, $\expval{ (\delta) (\epsilon)(\epsilon f \mu^2\theta)}=f\mu^2 \expval{\epsilon^2}\expval{\delta^2}$ and $\expval{(\epsilon)^3}$ sourced by short wavelength contributions $\epsilon$, $\epsilon\delta$ and $\epsilon f\mu^2\theta$ with $\epsilon$ a stochastic field uncorrelated with $\delta$ \cite{Ebina24,Ivanov22a}. As each of these quantities has an order of magnitude set by Poisson statistics, one expects $B_\mathrm{shot}$, $A_\mathrm{shot}=\mathcal{O}(1)$ and $P_\mathrm{shot}=\bar{n}N-1 = \mathcal{O}(1)$ \cite{Maus24b,Bakx25,Chudaykin25a}.

In addition, we include a contribution that is formally higher-loop order following ref.~\cite{Ebina24}. In typical EFT modeling, contributions from higher-loop orders are introduced only at small scales and hence by limiting the analysis to large scales, one can terminate modeling at finite order. For this work, we limit $M$ to (quasi-)linear scales ($k<0.12\,h^{-1}\,\mathrm{Mpc}$) and thus will not require typical EFT correction terms (counterterms)\footnote{If one aims to model smaller scales, one will require counterterm corrections, such as this ansatz used in ref.~\cite{Ebina24}
\begin{equation}
    M(k,\mu) \supset C_{\delta_M}(k) k^2 \left(N_{m,2}^{(0)} + \mu^2 N_{m,2}^{(2)} \right)
\end{equation}
where $C_{\delta_M}(k)=C_0 + C_1 W_R(k)$.
}. However, for the MPS there is a large-scale, perturbative\footnote{These corrections are similar to that in the Lyman-$\alpha$ forest two-point function, although with the critical distinction that the ones in MPS are UV-safe \cite{Ebina24}.}, higher-loop correction that arise from contact terms (fields at the same point in configuration space) \cite{Philcox21,Ebina24}. For this work, we only consider the MPS monopole and hence require only one free parameter 
\begin{equation}
    M_0 \supset A_0 P_L
\end{equation}
where, due to the perturbative nature of the correction, we estimate $A_{0}=\mathcal{O}(\sigma_R^4,\sigma_R^2\bar{\sigma}_R^2, \bar{\sigma}_R^4)\lesssim \sigma_R^2,\,\bar\sigma_R^2$ for the expansion coefficients defined as
\begin{align} \label{sigma_eqn}
    \sigma_R^2 = \int_{\bm{p}} W_R(p) P_L(p), &\qquad 
    \bar{\sigma}_{R}^2 = \int_{\bm{p}} (pR)^2 W_R(p) P_L(p), 
\end{align}
which, for the smoothing scales that we consider in this work, are $\lesssim 0.2$, 0.11, 0.07, for $R=10$, 15, and 20$\,h^{-1}\,\mathrm{Mpc}$ respectively, at $z=0.7$\footnote{The maximum correction (at $R=10\,h^{-1}\,\mathrm{Mpc}$) grows from $\lesssim0.15$ at $z=1.1$ \cite{Ebina24} to $\lesssim0.2$ at $z=0.7$.} \cite{Ebina24}. In practice, we find that we do not need to vary $A_0$ for any of our theory fits to mock catalogs in this work. However, we will continue to include $A_0$ in the discussion of the text following theoretical predictions. 
Thus in what follows we will consider only the following new nuisance parameters
\begin{equation}
    \{B_\mathrm{shot},A_\mathrm{shot}, A_{0} \}
\end{equation}
in addition to those from the one-loop power spectrum
\begin{equation}
    \{b_1, b_2, b_s,\alpha_0,\alpha_2,N,N_2\}
\end{equation}
and show that this provides the level of accuracy expected of next-generation surveys to jointly fit $P_0$, $P_2$, and $M_0$\footnote{We neglect the third order bias $b_3$ following standard practice \cite{DESI24-V} for power spectrum analyses, as $M$ does not introduce new $b_3$ dependencies.}. Note that choose to focus on the MPS monopole in this work due to modeling difficulties~\cite{Ebina24}, although the quadrupole and higher multipoles also have the potential to assist degeneracy breaking. This investigation will be left for future work, but detecting the value of $b_s$, which is a subdominant contribution to $M_0$ as described below, may be a potential use case. 

\subsection{Degeneracy Breaking}

The MPS depends upon bias parameters differently than the power spectrum, offering the possibility of breaking degeneracies. This is most apparent for the quadratic biases, $b_2$ and $b_s$, where the parameter dependence comes at leading order for $M$, in contrast to $P$ where they arise at one-loop. The $b_2$-degeneracy breaking power of the bispectrum has long been known to the community \cite{Bernardeau97,Pires12,Hahn21,Ivanov24} and has been confirmed for the marked spectrum in ref.~\cite{Ebina24}. Here, we will develop this insight further. 

The leading order contribution to (the bispectrum term in) the MPS comes in three contractions
\begin{equation}
    \expval{\delta_{g,R}^{(1)}(\mathbf{x_1})\delta_g^{(2)}(\mathbf{x_1})\delta_g^{(1)}(\mathbf{x_2})}, \quad \expval{\delta_{g,R}^{(2)}(\mathbf{x_1})\delta_g^{(1)}(\mathbf{x_1})\delta_g^{(1)}(\mathbf{x_2})}, \quad \expval{\delta_{g,R}^{(1)}(\mathbf{x_1})\delta_g^{(1)}(\mathbf{x_1})\delta_g^{(2)}(\mathbf{x_2})}
\end{equation}
which are different integrations over $B_{112}$, as shown in \S\ref{sec:PTintro}. 

Let us start with the first term which, following Eqn.~\ref{eqn:deltagn} for $\delta^{(n)}$, can be expanded as 
\begin{equation}
    \expval{\delta_{g,R,1}^{(1)}\delta_{g,1}^{(2)}\delta_{g,2}^{(1)}}_{\rm FT} = \int_\mathbf{p} Z_2(\mathbf{k},\mathbf{p}) Z_1(\mathbf{k}) Z_1(\mathbf{p}) W_R(p)P_L(k)P_L(p) 
\end{equation}
where we have abbreviated the coordinates to subscripts on the fields. The only $b_2$ dependence in this term is that sourced by $Z_2 \supset b_2/2$, yielding
\begin{align}
     \expval{\delta_{g,R,1}^{(1)}\delta_{g,1}^{(2)}\delta_{g,2}^{(1)}}_{\rm FT} &\supset Z_1(\mathbf{k}) P_L(k)\ \frac{b_2}{2}\int_\mathbf{p} Z_1(\mathbf{p}) W_R(p)P_L(p) \\
     &= \frac{b_2}{2} \left(b_1+\frac{f}{3}\right) Z_1(\mathbf{k}) P_L(k) \int \frac{p^2\,dp}{2\pi^2} W_R(p) P_L(p) 
     \label{eqn:b2eqn1}
\end{align}
which again highlights the importance of the smoothing, $W_R$, as the expression would otherwise be UV-sensitive. 

The second contribution $\expval{\delta_{g,R,1}^{(2)}\delta_{g,1}^{(1)}\delta_{g,2}^{(1)}}$ merely differs from the derivation above by the argument of the smoothing, which will now be $W_R(|\mathbf{k}+\mathbf{p}|)$. The third contribution and its $b_2$ dependence can be expanded as 
\begin{align}
 \expval{\delta_{g,R,1}^{(1)}\delta_{g,1}^{(1)}\delta_{g,2}^{(2)}}_{\rm FT} &= \int_\mathbf{p} Z_2(\mathbf{k}-\mathbf{p},\mathbf{p})Z_1(\mathbf{k}-\mathbf{p})Z_1(\mathbf{p}) W_R(p) P_L(|\mathbf{k}-\mathbf{p}|)P_L(p) \\
    &\supset \frac{b_2}{2} \int_\mathbf{p}  Z_1(\mathbf{k}-\mathbf{p})Z_1(\mathbf{p}) W_R(p) P_L(|\mathbf{k}-\mathbf{p}|)P_L(p)
\end{align}
which in the low-$k$ limit ($k\to 0$) contains
\begin{equation}
     \frac{b_2}{2} \left(b_1^2 + \frac{2b_1 f }{3} + \frac{f^2}{5} \right) \int \frac{p^2\,dp}{2\pi^2}W_R(p) P_L^2(p) 
\label{eqn:b2_in_112}
\end{equation}
having contributions $\propto b_2$, $b_1 b_2$, and $b_1^2 b_2$. These contributions are distinct from that in the power spectrum
\begin{align}
    P_{22} &= \expval{\left(\delta_g^{(2)}\right)^2}=2\int_\mathbf{p} Z_2^2(\mathbf{p},\mathbf{k}-\mathbf{p}) P_L(k)P_L(|\mathbf{k}-\mathbf{p}|) \\
    P_{13} &= \expval{\delta_g^{(1)}\delta_g^{(3)}}=3\int_\mathbf{p} Z_3(\mathbf{k},\mathbf{p},-\mathbf{p})Z_1(\mathbf{k}) P_L(k)P_L(p) 
\end{align}
which have a different set of kernels with different momenta as arguments. In particular, the $P_{22}$ term produces a $b_2^2$ contribution that cannot appear in the marked spectrum. Additionally, the $b_2$ dependence of $P$, which only enters at one-loop, is subdominant to linear theory terms, whereas in $M$ there is a $b_2$ dependence in every leading-order contraction. 

\begin{figure}[!ht]
    \centering
    \includegraphics[width=0.44\linewidth]{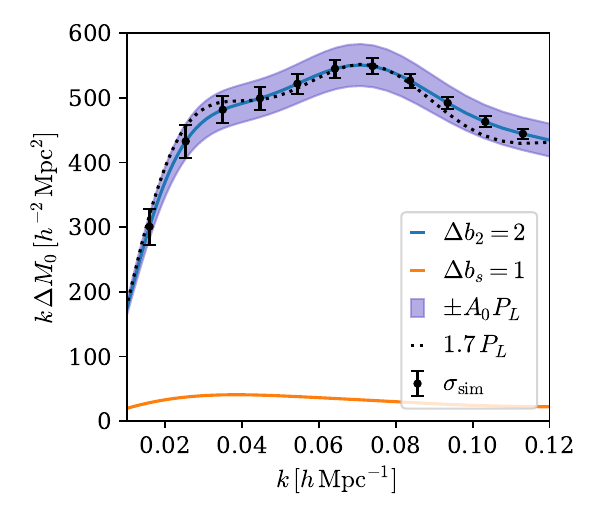}
    \includegraphics[width=0.55\linewidth]{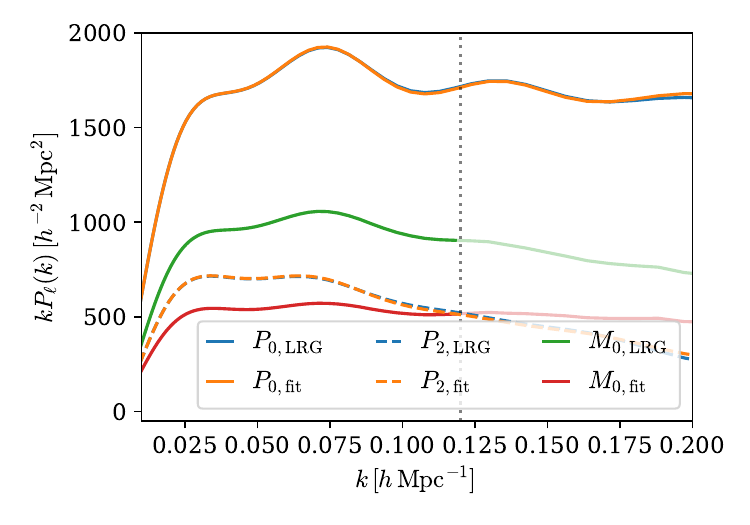}
    \caption{
    \textbf{Left:} The change in $M_0$ (with $R=15h^{-1}\,\mathrm{Mpc}$ at $z=0.7$) from shifts in second-order bias $b_2$ (blue line; $\Delta b_2 = 2$) and shear bias $b_s$ (orange; $\Delta b_s =1$) of size equal to the $1\,\sigma$ uncertainty from current generation surveys \cite{Chudaykin25a}. The $b_2$ dependence dominates despite both biases being quadratic in $\delta$. Overlaid on the $b_2$ curve is the error observed from $8\,h^{-3}\,\mathrm{Gpc}^3$ simulations (black error bars; \S\ref{sec:periodicbox}) and the anticipated maximum effect of higher-loop contributions (blue shade; \S\ref{sec:nuisance}) with $A_0 \lesssim0.11$, both of which are subdominant to the expected $b_2$ change. The dotted black line shows $1.7$ times the linear power spectrum $P_L$ demonstrating both that the shape of $\Delta M_0$ is degenerate with $P_L$ (Appendix \ref{sec:Mshape}) and that the higher-loop correction $A_0 P_L$ is subdominant by over an order of magnitude to the $b_2$ change. 
    \textbf{Right:} The power spectrum and $R=15h^{-1}\,\mathrm{Mpc}$ marked power spectrum of two nuisance parameter sets $\{\theta_\mathrm{LRG}\}$ and $\{\theta_\mathrm{fit}\}$ at $z=0.7$. $\theta_\mathrm{LRG}$ is the LRG-like bias parameters that fit the DESI DR1 cutsky mock at $z=0.7$ (\S\ref{sec:cutsky}) with $b_2\approx0.5$, while $\theta_\mathrm{fit}$ is another nuisance set with nearly identical power spectrum predictions but with $b_2=-1.5$. The similarity in $P_\ell(k)$ captures the $\Delta b_2\sim2$ fully expected in near-term redshift surveys, while $M_0(k)$ differs by a factor of 2, showing its potential to assist standard two-point analyses. The dotted vertical line indicates the $k_\mathrm{max}^M=0.12\,h^{-1}\,\mathrm{Mpc}$ used for this work.
    }
    \label{fig:degeneracy}
\end{figure}

Similarly, $b_s$ contributes to $Z_2$ at leading order and the $b_s$ dependence can be captured by substituting the $b_2$ term in $Z_2$ above by the $b_s$ term ($b_s [(\widehat{k}_1\cdot \widehat{k}_2)^2 - 1/3]$). For example, the $b_s$ dependence of the monopole of $\expval{\delta_{g,R,1}^{(1)}\delta_{g,1}^{(2)}\delta_{g,2}^{(1)}}$ is 
\begin{equation}
    \expval{\delta_{g,R,1}^{(1)}\delta_{g,1}^{(2)}\delta_{g,2}^{(1)}}_0 \supset\frac{8 f^2}{675} b_s P_L(k) \int \frac{p^2\,dp}{2\pi^2} W_R(p) P_L(p) 
\end{equation}
For $b_1=2$ and $f\sim0.9$, the monopole prefactor here is weaker than that of $b_2$ in Eqn.~\ref{eqn:b2eqn1} by a factor of $\sim550$, which indicates that using $M_0$ to detect $b_s$ may be difficult, as implied later in Fig.~\ref{fig:degeneracy}. The expression also indicates that this $b_s$ dependence is independent of the other nuisance parameters, but this is not a general result. For the second contraction $\expval{\delta_{g,R,1}^{(2)}\delta_{g,1}^{(1)}\delta_{g,2}^{(1)}}$
the `extra' exponential introduced by the change in smoothing argument ($W_R(p)\to W_R(|\mathbf{k}+\mathbf{p}|)$) breaks the result above, although it can be recovered in the low-$k$ limit. Finally the $b_s$ dependence of the third contraction is 
\begin{align}
    \expval{\delta_{g,R,1}^{(1)}\delta_{g,1}^{(1)}\delta_{g,2}^{(2)}}_{\rm FT} &\supset \int_\mathbf{p} b_s\left(\frac{((\mathbf{k}-\mathbf{p})\cdot\mathbf{p})^2}{|\mathbf{k}-\mathbf{p}|^2p^2}-\frac{1}{3}\right) Z_1(\mathbf{k}-\mathbf{p})Z_1(\mathbf{p}) W_R(p)P_L(|\mathbf{k}-\mathbf{p}|)P_L(p)
\end{align}
which, as $k\to 0$, reduces to
\begin{equation}
     b_s \left( b_1^2 + \frac{2 b_1 f}{3} + \frac{f^2 }{5} \right) 
    \int \frac{p^2\,dp}{2\pi^2} W_R(p) P_L^2(p) 
\end{equation}
taking a similar form to the dependency of $b_2$ in the same term (Eq.~\ref{eqn:b2_in_112}). 

Figure \ref{fig:degeneracy} shows the potential of $M_0$ for breaking degeneracy in practice for a tracer with LRG-like biases ($b_1=2.07$, $b_2=0.43$, $b_s=1.18$; fit to DESI DR1 cutsky mocks \S\ref{sec:cutsky}) at $z=0.7$. The left panel shows the effect of a $\sim 1\sigma$ deviation of $b_2$ and $b_s$ in current and near-future surveys ($\Delta b_2=2$, $\Delta b_s=1$; see e.g.~Table VII of ref.~\cite{Chudaykin25a}). We first observe that, as expected from preliminary results in the calculation above, the $b_2$ dependence is considerably stronger than that of $b_s$, indicating that $b_2$ will likely be the nuisance parameter of interest when employing the MPS, although the inclusion of the quadrupole $M_2$ may assist detecting $b_s$ deviations as well as other angular bispectrum signals.
The figure further shows that this `signal' dominates against both observational uncertainty ($\sigma_\mathrm{sim}$) and model uncertainty from higher-order corrections ($\pm A_0 P_L$) both by over an order of magnitude, showing certain potential for employment. The latter, in particular, is crucial, as the scale-dependence of the signal is largely degenerate with $P_L(k)$ and would be challenging to exploit without control over $A_0$. While here we find that the scale-dependence of $\Delta M_0$ resembles that of $P_L$, we also find that $M_{13}$ has a similar scale-dependence. This may be of interest in the future when making theoretical insights and has been explored in Appendix \ref{sec:Mshape}.
Now for the right panel let us consider two sets of biases: the LRG-like bias above ($\theta_\mathrm{LRG}$) and another ($\theta_\mathrm{fit}$) that returns an indistinguishable power spectrum but with a $b_2$ lower by 2. While $\Delta b_2=2$ here as in the left panel, it is a priori unclear whether the strong $b_2$ dependency seen in the left panel will fully manifest itself under this circumstance as we have altered other nuisance parameters as well. The figure, however, demonstrates that this is not of significant concern as the amplitude of $M_0$ deviates by a factor of 2, while $P_0$ and $P_2$ are indistinguishable. 

The two panels also offer some insight into what the binning of $M$ in data should be, as long as the $b_2$ signal is the principle aim. Both panels show that the $b_2$ dependence is smooth and weakly scale-dependent. This indicates that the precise values of $k_\mathrm{min}$ and $k_\mathrm{max}$ are not important, as this would only marginally change the statistical constraining power and would not reveal new information. This also suggests that a coarse binning will suffice, which is beneficial in terms of the covariance matrix measurement. 

\section{Covariance Matrix}
\label{sec:covariance}

To combine the power spectrum multipoles and the MPS we require not only an estimate of the covariance of each set of spectra, but also their cross-covariance. As the MPS has a significantly smaller data vector size than the bispectrum, we expect it will be possible to estimate the total covariance matrix between $P_\ell$ and $M_0$ using simulations, as was done for the recent power spectrum analysis of DESI DR1 full-shape data \cite{DESI24-V}. Regardless, it will be useful to gain analytic insight into the covariance structure. To this end, here we will discuss the covariance structure in real-space and neglect the effects of survey geometry for brevity, although extending to redshift-space will be straightforward in principle as described briefly in Appendix \ref{app:covariance}.

We start by defining the $N$-point correlator
\begin{equation}
    \expval{\delta(\mathbf{k}_1)\dots\delta(\mathbf{k}_N)} 
    = (2\pi)^3\delta_D\left(\sum_i \mathbf{k}_i\right) P_N(\mathbf{k}_1,\dots,\mathbf{k}_N)
\end{equation}
where $\delta_D$ is the Dirac delta function. 
As the following discussion will be centered around discrete $k$-bins with width $\Delta k$, it will also be useful to define the mode-counting factor, i.e.~the volume of each (thin) shell
\begin{equation}
    V_k = \int_{V_k} d^3 q\simeq 4\pi \int_{k-\Delta k/2}^{k +\Delta k/2} q^2 dq \approx 4\pi k^2 \Delta k
\end{equation}
and the translation between continuous and discrete $\delta$-functions
\begin{equation}
    (2\pi)^3 \delta_D(\mathbf{k}_1 - \mathbf{k}_2) \leftrightarrow V \delta^K_{k_1,k_2}
\end{equation}
where $V$ is the survey volume and $\delta^K$ is the Kronecker delta function. Using these definitions, the estimators for $P$ and $M$ become
\begin{equation}
    \widehat{P}_\mathbf{k} = V^{-1} \int_{V_{k}} \frac{d^3q}{V_k}\, \delta(-\mathbf{q})\delta(\mathbf{q}),  \qquad 
    \widehat{M}_\mathbf{k} = V^{-1} \int_{V_{k}} \frac{d^3q}{V_k}\, \delta(-\mathbf{q})\tilde\delta_M(\mathbf{q})
\end{equation}
where we have defined $\tilde{\delta}_M(\mathbf{q})=[\delta_{g,R}\star\delta_g](\mathbf{q})$ as a shorthand for the relevant contribution to the Fourier transform of $\delta_M$. Throughout this section we will also utilize the abbreviation $f(\mathbf{k})=f_\mathbf{k}$ to simplify the expressions. 

\subsection{Covariance of $M$}
\label{sec:autocov}

The covariance of a bin-averaged marked spectrum can be calculated as 
\begin{equation}
    \mathrm{Cov}\left(\widehat{M}_k, \widehat{M}_{k'}\right) = \expval{\widehat{M}_k \widehat{M}_{k'}} - \expval{\widehat{M}_k}\expval{ \widehat{M}_{k'}}   .
\end{equation}
The general form of this auto-covariance is largely similar to that for the filtered-square bispectrum \cite{Harscouet24}, although the smoothing kernel and the number of smoothed fields are different. The latter will appear as an increase in the number of terms due to a loss of symmetry between the two fields used to construct the composite field (as $\tilde\delta_M=\delta_{g,R}\delta_g$ and $\delta_{g,R}\neq\delta_g$ in this work). We will calculate the cross-covariance between $P$ and $M$ in \S\ref{sec:crosscov} after the auto-covariance.

Expanding the expression above we obtain
\begin{align}
    \mathrm{Cov}\left(\widehat{M}_k, \widehat{M}_{k'}\right) &= 
    V^{-2} \int_{V_k} \frac{d^3q}{V_k}  \int_{V_{k'}} \frac{d^3q'}{V_{k'}} \int_\mathbf{p} \int_{\mathbf{p}'} W_R(p) W_R(p') \\
    &\qquad (\left\langle \delta_{-\mathbf{q}} \delta_{\mathbf{p}} \delta_{\mathbf{q}-\mathbf{p}} \delta_{-\mathbf{q}'} \delta_{\mathbf{p}'} \delta_{\mathbf{q}'-\mathbf{p}'}\right\rangle 
    - \left\langle\delta_{-\mathbf{q}} \delta_{\mathbf{p}} \delta_{\mathbf{q}-\mathbf{p}}\right\rangle 
    \left\langle\delta_{-\mathbf{q}'} \delta_{\mathbf{p}'} \delta_{\mathbf{q}'-\mathbf{p}'} \right\rangle)
\end{align}

To calculate the whole covariance, we must consider all possible contractions in the six-field correlator. As each $\delta$ is a mean-zero field, contractions with $\expval{\delta}$ can be excluded. This leaves us with four possible categories of contractions: $(2,2,2)$ (e.g.~$\expval{ab}\expval{cd}\expval{ef}$), $(3,3)$ (e.g.~$\expval{abc}\expval{def}$), $(4,2)$ (e.g.~$\expval{abcd}\expval{ef}$), and $(6)$ (e.g.~$\expval{abcdef}$). These terms are traditionally referred to as $PPP$, $BB$, $PT$, and $P_6$, respectively, in bispectrum covariance calculations \cite{Biagetti22}. We run through the contractions for each category in Appendix \ref{app:covariance}. 

The calculations reveal that the covariance can be organized into diagonal ($D_{MM}$) and non-diagonal ($N_{MM}$) terms in the following fashion
\begin{align}
    \mathrm{Cov}\left(\widehat{M}_k, \widehat{M}_{k'}\right) &= (2\pi)^3 \delta^K_{k,k'} V^{-1} V_k^{-1} D_{MM} + V^{-1}N_{MM} \quad .
\end{align}
The inverse factor of $V_k\propto k^2$ accompanying $D_{MM}$ determines that it dominates at large-scales, where we have the best theoretical control. 
The diagonal term can be further factorized into each contraction type ($PPP$, $BB$, $PT$)
\begin{equation}
    D_{MM} = D_{222} + D_{33} + D_{42}
\end{equation}
where 
\begin{align}
    D_{222}(\mathbf{k}) &= \int_\mathbf{p} W_R(p)\big[ W_R(p) + W_R(|\mathbf{k}-\mathbf{p}|) \big] P(\mathbf{k})P(\mathbf{p})P(\mathbf{k}-\mathbf{p}) 
    \label{eqn:D222} \\    
    D_{33}(\mathbf{k}) &= M^2(\mathbf{k}) \\
    D_{42}(\mathbf{k}) &= \int_\mathbf{p} \int_{\mathbf{p}'} W_R(p)W_R(p')P(\mathbf{k})
    T(\mathbf{p},\mathbf{k}-\mathbf{p},\mathbf{p}',-\mathbf{k}-\mathbf{p}')
\end{align}
This full diagonal contribution can be recovered from the Gaussian (disconnected) covariance, i.e.
\begin{align}
    \mathrm{Cov}\left(\widehat{M}_k,\widehat{M}_k'\right) \supset \mathrm{Cov}_G\left(\widehat{M}_k, \widehat{M}_{k'}\right) = (2\pi)^3 \delta^K_{k,k'} V^{-1} V_k^{-1} \left[ D_{222}+D_{33}+D_{42} \right] 
\end{align}
indicating that for large-scale covariances the Gaussian approximation is sufficient. This behavior is anticipated and has been demonstrated previously for the filtered-square bispectrum covariance \cite{Harscouet24}; we further confirm this agreement along the covariance diagonal using the 25 periodic box mocks employed in this work (\S\ref{sec:periodicbox}). We note, however, that for the skew spectrum of type $\langle[\delta_R^2]\delta\rangle$, ref.~\cite{Hou23a} reports an appreciable SNR deviation of $\lesssim 50\%$ when the covariance is restricted to its diagonal entries. A precision study of the MPS covariance requires a larger simulation suite than is available here and therefore lies beyond the scope of this work, but revisiting this question in the future will be important.

\subsection{Covariance of $P$-$M$}
\label{sec:crosscov}

Now let us shift our attention to the covariance between $P$ and $M$,
\begin{align}
    \mathrm{Cov}\left(\widehat{P}_k, \widehat{M}_{k'}\right) &= 
    V^{-2} \int_{V_k} \frac{d^3k}{V_k}  \int_{V_k} \frac{d^3k'}{V_k} \int_{\mathbf{p}'} W_R(p') \\
    &\qquad (\left\langle \delta_{-\mathbf{q}} \delta_{\mathbf{q}} \delta_{-\mathbf{q}'} \delta_{\mathbf{p}'} \delta_{\mathbf{q}'-\mathbf{p}'}\right\rangle 
    - \left\langle\delta_{-\mathbf{q}} \delta_{\mathbf{q}}\right\rangle 
    \left\langle\delta_{-\mathbf{q}'} \delta_{\mathbf{p}'} \delta_{\mathbf{q}'-\mathbf{p}'} \right\rangle) 
\end{align}
With five density field operators, the possible contractions are of the form $(3,2)$ or $(5)$. 
While we refer the reader to Appendix \ref{app:covariance} for a detailed calculation, we once again find a factorization of covariance into diagonal ($D_{PM}$) and non-diagonal ($N_{PN}$) terms
\begin{align}
    \mathrm{Cov}\left(\widehat{P}_k, \widehat{M}_{k'}\right) &= (2\pi)^3 \delta^K_{k,k'} V^{-1} V_k^{-1} D_{PM} + V^{-1} N_{PM} \quad .
\end{align}
The inverse mode-counting factor, $V_k^{-1}\propto k^{-2}$, again indicates that $D_{PM}$ dominates at large scales.  This diagonal term only consists of contributions from the (3,2) contraction
\begin{equation}
    D_{PM}(\mathbf{k}) = D_{23} (\mathbf{k}) = 2 P(\mathbf{k}) M(\mathbf{k}) \quad .
    \label{eqn:D23}
\end{equation}
Similar to the covariance, we find that the Gaussian approximation recovers the full diagonal contribution, i.e.
\begin{align}
    \mathrm{Cov}_G\left(\widehat{P}_k, \widehat{M}_{k'}\right) 
    &= (2\pi)^3\delta^K_{k,k'} V^{-1} V_k^{-1} D_{23}
\end{align}
Due to the similarity in structure, this result carries over to other two-field composite fields (e.g.~the filtered square bispectrum) as well by a substitution of smoothing kernels.

\subsection{Scaling and relation to the bispectrum}

For the bispectrum, the cross-covariance is known to be subdominant to the auto-covariances for general triangles with sides of similar length ($k_1\sim k_2\sim k_3$), becoming important only for particular configurations such as the squeezed bispectrum ($k_1\ll k_2\simeq k_3$) \cite{Biagetti22,Salvalaggio24}. However, recent work shows that the correlations between the monopoles of the power spectrum and bispectrum are large $\left(\gtrsim0.5\right)$ at quasi-linear scales ($k\sim0.1\,h\,\mathrm{Mpc}^{-1}$) suggesting that the cross-covariance is not negligible for all compressions of the bispectrum \cite{Bansal26}.
If the cross-covariance is subdominant, it is both a convenient and scientifically motivating feature, as a smaller cross-covariance decreases the uncertainty requirement on measurement and covariance matrix, and qualitatively indicates that there is more `new' information in the probe. 
We will investigate how this situation translates to the MPS. 

The small cross-covariance of the bispectrum can be traced back to the difference in the mode-counting factor normalizing the estimators. For bins of width $\Delta k\ll k_i$, the volume of triangles included in the bispectrum is
\begin{align}
    V_\mathrm{tr} \simeq \int_{\mathbf{q}_1\in k_1} d^3q_1 \int_{\mathbf{q}_2\in k_2} d^3q_2 \int_{\mathbf{q}_3\in k_3} d^3q_3\ \delta_D(\mathbf{q}_{123}) \simeq 8\pi^2 k_1 k_2 k_3 \Delta k^3
\end{align}
where $\int_{\mathbf{q}\in k}d^3q = \int d\Omega_q \int_{k-\Delta k/2}^{k+\Delta k /2} q^2 dq$ and $\mathbf{q}_{123}=\mathbf{q}_1+\mathbf{q}_2+\mathbf{q}_3$ \cite{Scoccimarro97,Biagetti22}. Similar to the power spectrum, where $\widehat{P}\propto V_k^{-1}$, the bispectrum estimator scales as $\widehat{B}\propto V_\mathrm{tr}^{-1}$. 

This mode-counting factor enters the (auto-)covariance as 
\begin{align}
    \mathrm{Cov}\left[\widehat{B}(t),\widehat{B}(t')\right] &\simeq \mathrm{Cov}\left[\widehat{B}(t),\widehat{B}(t')\right]_{222}  
    \propto V_\mathrm{tr}^{-2} V_\mathrm{tr} P(k_1)P(k_2) P(k_3) 
\end{align}
where $V_\mathrm{tr}^{-2}$ comes from the two powers of $\widehat{B}$ and the $V_\mathrm{tr}$ from the integrals present after performing momentum-matching using $\delta_D$'s \cite{Biagetti22,Salvalaggio24}. 
This is distinct from the cross-covariance 
\begin{align}
    \mathrm{Cov}\left[\widehat{P}(k),\widehat{B}(t')\right] &\simeq \mathrm{Cov}\left[\widehat{P}(k),\widehat{B}(t')\right]_{32} 
    \propto V_\mathrm{tr}^{-1}V_k^{-1} V_\mathrm{tr} P(k) B(k_1,k_2,k_3) 
\end{align}
where the estimators $\widehat{P}$ and $\widehat{B}$ contribute $V_k^{-1}$ and $V_\mathrm{tr}^{-1}$, respectively, and the integrals after momentum-matching contribute $V_\mathrm{tr}$ \cite{Salvalaggio24}. This results in an overall volume factor of
$V_k^{-1}$, which is distinct from the auto-covariance scaling as $V_\mathrm{tr}^{-1}$. This is the key reason for the cross-covariance suppression of the bispectrum, as 
\begin{equation}
    \frac{\mathrm{Cov}\left[\widehat{P},\widehat{B}\right]}{\sqrt{\mathrm{Cov}\left[\widehat{P},\widehat{P}\right]
    \mathrm{Cov}\left[\widehat{B},\widehat{B}\right]}} 
    \sim \frac{V_k^{-1} P B}{\sqrt{V_k^{-1} V_\mathrm{tr}^{-1} P^5} }
    \sim \frac{\Delta k}{k} \sqrt{k^3 P}
\end{equation}
where we assume $k_1\sim k_2\sim k_3\sim k$ and approximate $B(k,k,k)\sim P^2(k)$, and $k^3P$ is dimensionless \cite{Biagetti22}. Hence, when $\Delta k\ll k$, the cross-covariance is suppressed by $\Delta k/k$. 
Since this suppression is a consequence of the difference in the mode-counting factor, the cross-covariance is `unsuppressed' as one considers larger $k$-bins (e.g.~$\Delta k \sim k$) and allow more triangles configurations to contribute, indicating that it may be misguided to use the suppression of cross-covariance per bin as a heuristic of new scientific information. In what follows, we will see an analogous situation for the MPS.

To trace the mode-counting factor of the MPS, let us define 
\begin{align}
    V_M &= \int_{\mathbf{q}\in k} \int_{\mathbf{p}_1} \int_{\mathbf{p}_2} W_R(p_1) \delta_D(\mathbf{q}+\mathbf{p}_1+ \mathbf{p}_2)
\end{align}
For the purpose of this exercise, it will also be useful to define a narrow smoothing kernel\footnote{For example, one could imagine constructing this by using as a mark the difference between two densities with similar but unequal smoothing lengths.} 
\begin{align}
    W_{\mathrm{hat},p_*}(k) = 
    \begin{cases}
        1 & p_*-\frac{\Delta p}{2} <k<p_*+\frac{\Delta p}{2}\\
        0 & \mathrm{otherwise}
    \end{cases}
\end{align}
where the mode-counting factor becomes $ V_M \approx 16 \pi^2 k^2 p_*^2 \Delta k \Delta p $. In what follows we will use this narrow kernel when explicitly invoking the kernel to make the mode-counting factor more transparent. In terms of mode-counting, the broad Gaussian kernel used for this work can roughly be recovered when $\Delta p \sim p_*$.

Let us now consider the dominant diagonal contribution to the cross-covariance $D_{23}$ (Eqn.~\ref{eqn:D23})
\begin{align}
    (2\pi)^3 \delta^K_{k,k'} V^{-1} V_k^{-1} D_{23} 
    &= (2\pi)^3 V^{-1} V_k^{-1} \delta^K_{k,k'} P(\mathbf{k}) M(\mathbf{k}) \\
    &= (2\pi)^3 V^{-1} V_k^{-1} \delta^K_{k,k'} P(\mathbf{k}) \int_\mathbf{p} W_R(p) B(\mathbf{p},\mathbf{k}-\mathbf{p},-\mathbf{k}) \\
    &= V^{-1} V_k^{-2} V_M \delta^K_{k,k'} P(\mathbf{k}) \langle B(\mathbf{p}_*,\mathbf{k}-\mathbf{p}_*,-\mathbf{k}) \rangle_\varphi
\end{align}
where at the final step we apply the thin-shell approximation $B(\mathbf{p}',\mathbf{q}-\mathbf{p}',-\mathbf{q})=B(\mathbf{p}_*,\mathbf{k}-\mathbf{p}_*,-\mathbf{k})$.
Similarly for the auto-covariance, the (2,2,2) diagonal term $D_{222}$ (Eqn.~\ref{eqn:D222}) scales as
\begin{align}
    & (2\pi)^3 \delta^K_{k,k'} V^{-1} V_k^{-1} D_{222} \nonumber \\
    &= (2\pi)^3 \delta^K_{k,k'} V^{-1} V_k^{-1} \int_\mathbf{p} W_R(p)\big[ W_R(p) + W_R(|\mathbf{k}-\mathbf{p}|) \big] P(\mathbf{k})P(\mathbf{p})P(\mathbf{k}-\mathbf{p}) \\
    &\approx V^{-1} V_k^{-2}V_M \delta^K_{k,k'} P(\mathbf{k})P(\mathbf{p}_*) \langle P(\mathbf{k}-\mathbf{p}_*) \rangle_\varphi
\end{align}
Note that we have not shown that the (2,2,2) term dominates out of the three diagonal terms, but we consider it to draw parallels with the bispectrum case. 

Thus, if we adopt the approximation $p_*\sim k\sim|\mathbf{k}-\mathbf{p}_*|$ and (crudely) approximate a constant contribution over the angle average $\langle f \rangle_\varphi$, the ratio between the cross- and auto-covariances scale as
\begin{equation}
    \frac{\mathrm{Cov}\left[\widehat{P},\widehat{M}\right]}{\sqrt{\mathrm{Cov}\left[\widehat{P},\widehat{P}\right]
    \mathrm{Cov}\left[\widehat{M},\widehat{M}\right]}} 
    \sim \frac{V_k^{-2} V_M B}{\sqrt{V_k^{-3} V_M P^5}}
    \sim \frac{V_M^{1/2} P^{1/2}}{V_k^{1/2}}
    = \sqrt{\frac{k^2 p_*^2 \Delta k \Delta p \, P}{k^2 \Delta k}}
    = \frac{p_*}{k}\sqrt{\frac{\Delta p}{k}} \sqrt{k^3 P}
\end{equation}
showing a suppression of the cross-covariance by $\sqrt{\Delta p/k}$ as opposed to the $\Delta k/k$ of the bispectrum. 
By extension, we see that the MPS considered in this work, with a Gaussian smoothing kernel $W_R$, corresponds to the case $\Delta p \sim k$, with little suppression. 
This is a direct consequence of the change in mode-counting factors. 
Simply put, the (wide) integration over the inner momentum $\mathbf{p}$ in $M$ means that there are far more triangles that contribute, whereas for $B$ only specific triangle configurations give matches. This is analogous to the suppression in $P-B$ cross-covariance becoming weaker when considering wider $k$-bins and allowing more triangles to contribute.  
Note that a wide window breaks the thin-shell approximation applied above and thus is merely an analogy valid for order-of-magnitude estimates. 
This result is in qualitative agreement with past work investigating the cross-covariance of skew spectra with the power spectrum, which found weaker but significant correlations \cite{Dai20,Hou23a}.

A stronger suppression with a narrow window indicates that there is more ``new'' information when using a narrow window and will be less affected from data uncertainties when computing the inverse covariance. 
Of course, this is in trade-off with other factors influenced by the window choice, such as the SNR, degeneracy breaking in the whole parameter space, and any numerical uncertainties and systematics dependence when using sharp windows with significant ringing in real-space. 
In particular, it is expected that the SNR will be impacted, as one is decreasing the number of data points that contribute. 

A full investigation of the results above is outside of the scope of this work, as it will require a large suite of simulations considering specific survey footprints. At the present stage, we still conduct a preliminary validation of the results using 25 mock catalogs in $2\,h^{-1}\,\mathrm{Gpc}$ periodic boxes, which are later introduced in \S\ref{sec:periodicbox}. We indeed find that the Gaussian contribution dominates and the cross-covariance is not significantly suppressed relative to the auto-covariances, indicating that we are in a qualitatively different situation than when considering the joint analyses of $P$ and $B$.

\section{The Effect of Survey Geometry}
\label{sec:window}

Previous calculations and applications of the MPS were limited to those on periodic boxes\footnote{Ref.~\cite{Cowell25} has applied the marked angular power spectrum to weak lensing data, but did not model the window effect as it used an emulator over simulations as the theoretical prediction.}. However, real galaxy surveys observe only a fraction of the sky, requiring an estimator that accounts for the (often complex) survey footprint. This technique is established for power spectrum calculations and we demonstrate how this can be carried over directly to the MPS. 

\subsection{Power Spectrum}

The modeling of the power spectrum over a part of the sky is well-established \cite{FKP}. Observationally, the galaxy distribution is captured by the weighted galaxy density, $n_g$, and it is compared to a weighted random density, $n_r$, which captures the survey geometry and mean density variation without any clustering. The randoms thus encode information about the survey footprint, which can be defined as a survey geometry selection function $W(\mathbf{x})$ through ensemble averages over the two densities 
\begin{equation}
    W(\mathbf{x}) = \expval{n_g(\mathbf{x})} = \alpha\expval{n_r(\mathbf{x})}
    \quad\mathrm{with}\quad
    \alpha = \frac{\int d\mathbf{x} \ n_g(\mathbf{x})}{\int d\mathbf{x} \ n_r(\mathbf{x})}
    \quad .
\end{equation}
Using these ingredients, the power spectrum multipoles over the survey footprint can be estimated by \cite{Yamamoto05}
\begin{equation}
    \widehat{P}_\ell(k_\mu) = \frac{2\ell + 1}{AV_{k_\mu}} \int_{V_{k_\mu}} \mathrm{d}\mathbf{k} \int \mathrm{d}\mathbf{x}_1 \int \mathrm{d}\mathbf{x}_2 \, e^{i\mathbf{k} \cdot (\mathbf{x}_2 - \mathbf{x}_1)} \mathcal{F}(\mathbf{x}_1) \mathcal{F}(\mathbf{x}_2) \mathcal{L}_\ell(\widehat{\mathbf{k}} \cdot \widehat{\mathbf{x}}_1) - \mathcal{N}_\ell
\end{equation}
where $\mathcal{F}(\mathbf{x})=n_g(\mathbf{x}) - \alpha n_r(\mathbf{x})$ is the FKP field \cite{FKP}, $V_{k_\mu}$ is the volume of the shell of $\mathbf{k}$ being integrated over, $\mathcal{L}_\ell$ is the Legendre polynomial of order $\ell$ and $\mathcal{N}_\ell$ is the estimated shot-noise contribution.  The normalization, $A$, is conventionally taken to be\footnote{The product of $n_g$ and $n_r$ is used instead of the product of two $n_r$'s to avoid a shot-noise bias. See \url{https://pypower.readthedocs.io/en/latest/api/api.html\#pypower.fft_power.normalization} for a numerical implementation.}
\begin{equation}
    A = \alpha \int d\mathbf{x}\ n_g(\mathbf{x})n_r(\mathbf{x})
    \approx \int d\mathbf{x}\ \expval{n_g(\mathbf{x})}^2
\end{equation}
where $\expval{n_g(\mathbf{x})}$ corresponds to the mean density. 

\begin{figure}
    \centering
    \includegraphics[width=0.98\linewidth]{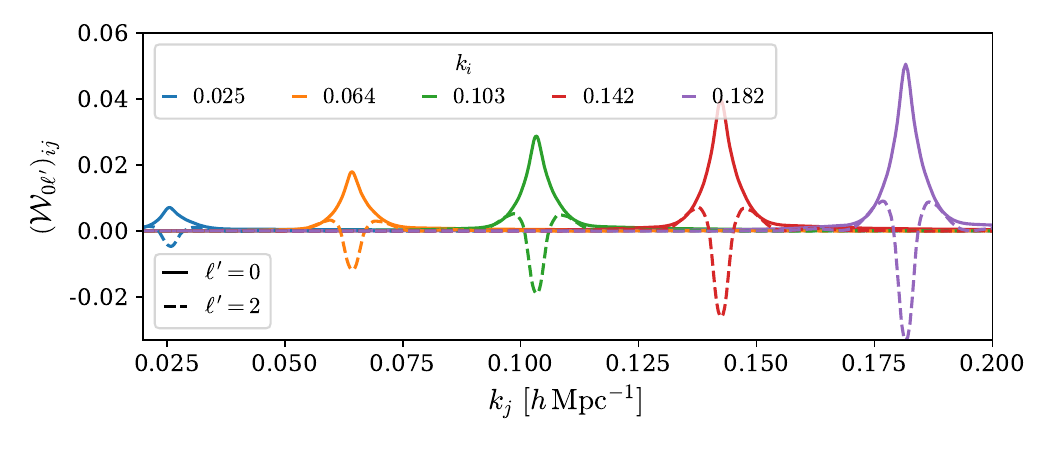}
    \caption{Some pieces of the Fourier-space DR1 cutsky window matrix $(\mathcal{W}_{\ell \ell'})_{ij}$ (Eqns.~\ref{eqn:windowP} and \ref{eqn:windowM}).  We choose $\ell=0$ and select $k_i$, for bins of width $\Delta k\sim0.01$. The real-space window matrix is shown in Fig.~6 of ref.~\cite{Chaussidon25}.}
    \label{fig:window}
\end{figure}

For Poisson shot-noise $\expval{\mathcal{F}(\mathbf{x})\mathcal{F}(\mathbf{x}')}=W(\mathbf{x})W(\mathbf{x}')\xi(\mathbf{x},\mathbf{x}') + W(\mathbf{x})\delta_D(\mathbf{x}-\mathbf{x}')$ \cite{FKP}.  The $\delta_D$ term is subtracted by $\mathcal{N}_\ell$ above and so for a thin bin in $k$ the expectation of $\widehat{P}_\ell$ becomes \cite{Castorina18,Beutler21}
\begin{align}
    \expval{\widehat{P}_{\ell}(k)} &= \frac{2\ell+1}{A} 
    \int \frac{d\Omega_{k}}{4\pi}
    \int d\mathbf{x}_{1} \int d\mathbf{x}_{2}\,
    e^{i\mathbf{k}\cdot(\mathbf{x}_{2}-\mathbf{x}_{1})}\,W(\mathbf{x}_{1})\,W(\mathbf{x}_{2})\,
    \mathcal{L}_{\ell}\!\left(\hat{\mathbf{k}}\cdot\hat{\mathbf{x}}_{1}\right)\,
    \xi(\mathbf{x}_{1},\mathbf{x}_{2})  \\
    &= \frac{2\ell+1}{A} \sum_{p}  \int \frac{d\Omega_{k}}{4\pi} \int d\mathbf{x}
    \int d\mathbf{s}\, e^{-i\mathbf{k}\cdot\mathbf{s}} W(\mathbf{x})W(\mathbf{x}-\mathbf{s})
    \mathcal{L}_{\ell}\!\left(\hat{\mathbf{k}}\cdot\hat{\mathbf{x}}\right)
    \mathcal{L}_p\!\left(\hat{\mathbf{x}}\cdot\hat{\mathbf{s}}\right) \xi_{p}(s)   \\
    &= 4\pi(-i)^{\ell}(2\ell+1) \sum_{\ell_{1},\ell_{2}}
    \begin{pmatrix}
    \ell_{1} & \ell_{2} & \ell \\
    0 & 0 & 0
    \end{pmatrix}^{2}
    \int s^2\,ds\   \xi_{\ell_{1}}(s)\,  \mathcal{W}_{\ell_{2}}(s)\, j_{\ell}(k s)
\end{align}
where we define $\mathbf{s}=\mathbf{x}_1-\mathbf{x}_2$ and the real space window matrix $\mathcal{W}$ is defined as 
\begin{equation}
    \mathcal{W}_\ell(s) = \frac{2\ell+1}{4\pi A} \int d\Omega_s \int d\mathbf{x}\  W(\mathbf{x}) W(\mathbf{x}-\mathbf{s})\mathcal{L}_\ell(\hat{\mathbf{x}}\cdot \hat{\mathbf{s}})
\end{equation}
While this does not include wide-angle corrections to the plane-parallel approximation, these can be incorporating by expanding the correlation function $\xi$ in powers of $s/d$, where $\mathbf{s}$ is the pair separation and $\mathbf{d}$ is the line-of-sight distance \cite{Castorina18,Beutler21}. The window matrix at each order then becomes 
\begin{equation}
    \mathcal{W}^{(n)} = \frac{2\ell+1}{4\pi A} \int d\Omega_s \int d\mathbf{x} 
    \ x^{-n} W(\mathbf{x}) W(\mathbf{x}-\mathbf{s})\mathcal{L}_\ell(\hat{\mathbf{x}}\cdot \hat{\mathbf{s}})
\end{equation}
Following ref.~\cite{Chaussidon25} we will include the first order correction ($n=1$). 
The convolved power spectrum over a finite $k$-bin, $k_i$, can then be modeled as a matrix multiplication of the window matrix and the theory model
\begin{equation}
    (\widehat{P}^\mathrm{obs}_\ell)_{i}  = (\mathcal{W}_{\ell \ell'})_{ij} (P_{\ell'})_j \quad .
    \label{eqn:windowP}
\end{equation}
We show in Fig.~\ref{fig:window} some slices of the window matrix, $\mathcal{W}$,
for the DESI DR1 cutsky simulations considered in \S\ref{sec:cutsky}.

\subsection{Marked Power Spectrum}

The estimation of the marked power spectrum, including the effects of the survey geometry, closely follows the power spectrum calculation. 
Using the expression for the overdensity field $\delta_g = (n_g - \alpha n_r)/ \alpha n_r$ one can describe the marked field as
\begin{align}
    \bar{m}\delta_M  = m (\delta_g+1)-\bar{m} = \frac{m n_g}{\alpha n_r}-\bar{m} = \frac{mn_g-\bar{m}\alpha n_r}{\alpha n_r} .
\end{align}
In the power spectrum calculation, the FKP field $\mathcal{F}$ is related to the overdensity field by a factor of the mean density
\begin{equation}
    \mathcal{F}(\mathbf{x})=n_g(\mathbf{x}) - \alpha n_r(\mathbf{x})
    \quad\mathrm{c.f.}\quad
    \delta_g = \frac{n_g - \alpha n_r}{\alpha\, n_r}
\end{equation}
In parallel to this, we define
\begin{align}
    \mathcal{F}_M(\mathbf{x})=m(\mathbf{x}) n_g(\mathbf{x})-\bar{m}\alpha\, n_r(\mathbf{x})
\end{align}
which substitutes $\mathcal{F}$ in the power spectrum estimator, while holding the remaining components the same, including the normalization $A$. Since $M$ is also a 2-point function and the geometrical effect is fully captured by the randoms designed for the power spectrum, the same window matrix ($\mathcal{W}$) established to model the power spectrum over a fraction of the sky can be reused, making the overall equation simply
\begin{equation}
    (\widehat{M}^\mathrm{obs}_\ell)_{i}  = (\mathcal{W}_{\ell \ell'})_{ij} (M_{\ell'})_j
    \label{eqn:windowM}
\end{equation}
This is a major advantage, as there is no need to rebuild a new window formalism incorporating the same corrections and it bypasses the need for many validations. 

The above derivation neglects the effect of the survey geometry on the mark, i.e.\ it assumes that $\delta_M^\mathrm{obs}(x)\approx m(x) W(x)\delta_g(x)$.
As long as the survey volume is much larger than $R^3$ and not full of holes this approximation is well-justified.  Phrased another way, the mark is a local function of the density field and the window effect is most significant at large scales. 
Furthermore, our use of this approximation is the same as that employed in the standard post-reconstruction BAO analysis  (with smoothing $R=15\,h^{-1}\,\mathrm{Mpc}$), which itself applies a (likely) less local operation on $\delta_g$ and yet yields results consistent with the configuration-space analysis that fully incorporates the window in the 2PCF. 

This approach to survey geometry differs from those in the past literature. Ref.~\cite{Chen24} approximates the window convolution of the skew spectrum by decomposing the tree-level bispectrum into products of power spectra and convolving each leg with the power spectrum window matrix. This approximation is fairly successful with few-percent errors typically, but can reach up to 18\% deviations for select shapes in the presence of primordial non-Gaussianity. Recently, ref.~\cite{deBelsunce25} has developed the exact window treatment for the Lyman-$\alpha$ forest skew spectrum $\langle\delta_F^2\,\delta_F\rangle$, where the absence of derivative operators in $\delta^2$ allows a clean factorization of the window from the three-point correlator, yielding a window kernel involving $W^2(\mathbf{x}_1)W(\mathbf{x}_2)$ that can be computed via pair counting. For the MPS or galaxy skew spectra involving (transversely) smoothed fields, this factorization is complicated by the smoothing kernel, which entangles the window with the density field on scale $R$, although the MPS minimizes this entanglement by involving only the single smoothing kernel required to regulate UV divergences in the quadratic field. 
For galaxy surveys, the dense three-dimensional sampling furthermore permits a robust estimate of the smoothed density field in the survey interior using random catalogs, in the same manner as BAO reconstruction, such that the mark is approximately independent of the survey geometry. Under this approximation the window treatment of the MPS reduces to that of a standard two-point function. We validate this on cutsky mocks in \S\ref{sec:cutsky}, finding sub-$\sigma$ agreement across all modes. The situation is fundamentally different for the Lyman-$\alpha$ forest, where the sparse transverse sampling of quasar sightlines makes a stable local estimate of the smoothed selection function impractical, necessitating the explicit forward modeling of the window.

\subsection{Implementation}

We implement the algorithm above largely inheriting the code structure of \texttt{pypower}\footnote{\url{https://github.com/cosmodesi/pypower}} \cite{Hand17}, which includes both the methods necessary for two-point correlator measurements and window matrix calculations. As mentioned above, the latter does not require modifications, all of the changes will be in the former; in particular, we will aim to replace the FKP field $\mathcal{F}$ with $\mathcal{F}_M=mn_g-\bar{m}\alpha n_r$.

To this end we require a robust way to measure the mark $m$, which is dependent on the measurement of the smoothed field $\delta_{g,R}$ over an inhomogeneous footprint. This is a well-known technique to the field, as $\delta_{g,R}$ is a key component of BAO reconstruction. For our implementation we will follow that of \texttt{pyrecon}\footnote{\url{https://github.com/cosmodesi/pyrecon}} 
\begin{equation}
    \delta_{g,R}(\mathbf{x})=\frac{n_{g,R}(\mathbf{x})-\alpha n_{r,R}(\mathbf{x})}{\alpha n_{r,R}(\mathbf{x})}
\end{equation}
where $n_{g,R}(\mathbf{k})=W_R(k) n_g(\mathbf{k})$ and $n_{r,R}(\mathbf{k})=W_R(k) n_r(\mathbf{k})$. For stability we set $\delta=0$ for mesh points with less than 1\% of the average random weight. Note that the density field calculation increases the importance of the density of randoms and mesh grid size compared to a standard power spectrum calculation. This will be revisited at the end of this section. 
Given this stable density calculation, measuring $\mathcal{M}$ (Eqn.~\ref{defMorig}) from data is straightforward\footnote{The software used for this study will be incorporated into the DESI pipeline in the future.}. 

From a practical point of view of a $P+M$ fit, it is beneficial to fit the higher-point information $M$ directly, instead of $\mathcal{M}$, which (doubly) includes two-point information from $P$, making the modeling of off-diagonal covariance more important. In order to measure $M$ there is one additional procedure necessary to account for the smoothing of two-point information. Since the smoothing $W_R(k)$ is not commutative with the effect of the survey geometry, 
\begin{equation}
    M_\mathrm{obs} \neq \mathcal{M}_\mathrm{obs}-(C_0^a+C_1^a W_R(k))(C_0^b+C_1^b W_R(k)) P_\mathrm{obs}
\end{equation}
Rather, we must consider the smoothed power spectra individually and subtract their contributions
\begin{equation}
    M_\mathrm{obs} = \mathcal{M}_\mathrm{obs}- (C_0^a C_0^b P_\mathrm{obs} + (C_0^a C_1^b + C_0^b C_1^a)P_{R,\mathrm{obs}} + C_1^a C_1^b P_{RR,\mathrm{obs}})
\end{equation}
where $P_R=\expval{\delta_{g,R}\delta_g}$ and $P_{RR}=\expval{\delta_{g,R}^2}$ can be computed analogously to $P$ itself. 

Although the necessity to measure the density field numerically increases the importance of the density of randoms and meshgrid size compared to a standard power spectrum measurement, we have verified that the algorithm performance converges within reasonable requirements. We validate the convergence of the code using both periodic box and `cutsky' simulations, as introduced later in \S\ref{sec:validation}. 
In this work we adopt algorithm settings of meshsize of $2\, h^{-1}\,\mathrm{Mpc}$ for the periodic box and $3.5\, h^{-1}\,\mathrm{Mpc}$ for cutsky, and a random density of $3.5\times10^{-2} \,h^3\,\mathrm{Mpc}^{-3}$ (all 18 random catalogs associated) with cutsky. With these settings, we achieve performance that converges to $\lesssim 1\%$ in MPS amplitude, which is sufficient for this work. We do not require randoms for the periodic box since we can assume uniform random density. 

\section{Validation on mocks}
\label{sec:validation}

\subsection{Periodic box mocks}
\label{sec:periodicbox}

Here we perform validations of the theory against mock catalogs generated on the periodic boxes from the \textsc{AbacusSummit} N-body simulation suite \cite{Maksimova21}, produced with the {\sc Abacus} N-body code \cite{Garrison18,Garrison21}. We adopt the 25 `base' (2$\,h^{-1}\,\mathrm{Gpc}$) boxes at $z=0.8$, with the DESI LRG2 tracer in mind ($0.6<z<0.8$), and populate the N-body simulation with halos generated using the standard, five-parameter Halo Occupation Distribution (HOD) model \cite{Zheng07} implemented in AbacusUtils\footnote{\url{https://abacusutils.readthedocs.io/en/latest/}} software \cite{Yuan22}. The standard HOD model defines probability distribution of central and satellite galaxies depending on host halo mass. The central and satellite galaxies are populated based on a binomial and Poisson-based distribution, respectively. 
This can be described by
\begin{align}
\langle N_{\text{cen}}(M_h) \rangle &= \frac{1}{2}  \text{erfc} \left( \frac{\ln{M_{\text{cut}}/ M_h}}{\sqrt{2}\sigma} \right)
\label{eqn:hod-cen} \\
\langle N_{\text{sat}}(M_h) \rangle &= \langle N_{\text{cen}}(M_h) \rangle \left( \frac{M_h - \kappa M_{\text{cut}}}{M_1} \right)^{\alpha}
\quad \text{for } M_h > \kappa M_{\text{cut}}
\label{eqn:hod-sat}
\end{align}
where $M_h$ is the halo mass and \{$M_{\rm cut}$, $M_1$, $\sigma$, $\kappa$, $\alpha$\} are model parameters. 

Using the best-fit parameters for the LRG2 sample in the DESI one-percent survey \{$\log M_{\mathrm{cut}}=12.78$, $\log M_{1}=13.94$, $\sigma=0.17$, $\alpha=1.07$, $\kappa=0.55$\} \cite{Yuan23}, we generate mock catalogs.  We omit the incompleteness fraction, $f_\mathrm{ic}$, from the HOD fit and simply randomly downsample the mock galaxies to produce mocks at two number densities, $\bar{n}=10^{-3}$ and $3\times10^{-4} \, h^{-3}\,\mathrm{Mpc}^3$, which roughly spans the range of DESI galaxy densities \cite{DESI24-IV,DESI-DR2}. The mock galaxies have linear bias $b_1\approx2.11$ \cite{Yuan23} and power spectrum monopoles and quadrupoles as shown in the blue and orange lines in Fig.~\ref{fig:fit_grid}. 

We will use the mean and standard deviation between the boxes to validate the MPS theory for near-future data. 
The volume of each box corresponds to the volume of the LRG2 $z$-bin ($0.6<z<0.8$) but with a sky area of $\approx22000\deg^2$. This is $\approx 30\%$ larger than the footprint at the end of the DESI survey ($17000\deg^2$), providing a sufficient theory validation for the data in the near future. 
This is in similar spirit to the effort in ref.~\cite{Ebina24}, but differs crucially in the (reduced) number of nuisance parameters, as discussed in \S\ref{sec:nuisance}, and the question of stochasticity of the marked field as discussed below.

\subsection{Stochasticity with low number density}
\label{sec:stochasticity}

It has recently been pointed out \cite{Karcher24} that the MPS involves a new stochasticity due to estimating the marked field $m$ (and hence density field $\delta_{g,R}$) from a finite number of objects. 
This problem can be restructured as an introduction of a new stochastic field $\epsilon'$ for the MPS that is distinct from the power spectrum. To the extent that this stochasticity is scale-independent, this would introduce a new nuisance parameter $N'$ where the power spectrum shot noise enters the MPS. However, as evident from Eqn.~\ref{eqn:stochastic}, the contribution is largely degenerate with $B_\mathrm{shot}$ when considering only the monopole $M_0$. By fitting to mock catalogs with varying number densities with the same non-stochastic nuisance parameters, we demonstrate that this is indeed the case practically. 

\begin{figure}
    \centering
    \includegraphics[width=0.98\linewidth]{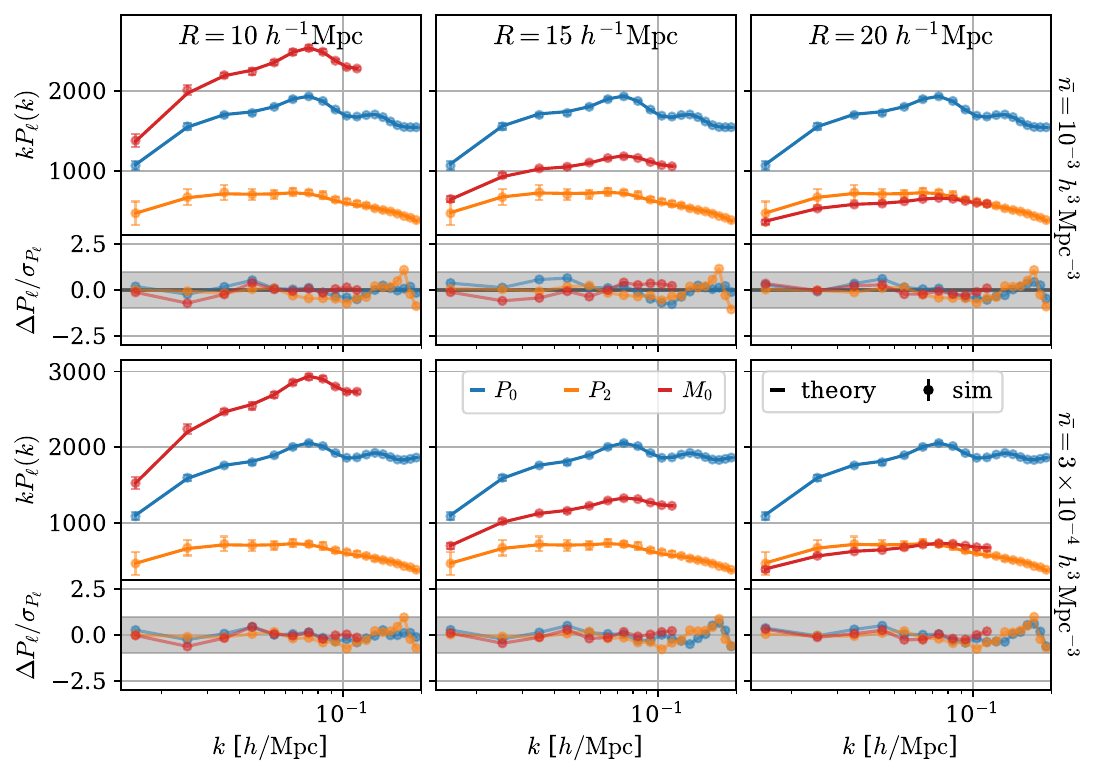}
    \caption{$P_\ell$ and $M_0$ joint-fits to the average of 25 simulations in $8\,h^{-3}\,\mathrm{Gpc}^3$ periodic boxes, with each column corresponding to smoothing scales $R=10$, 15, $20\, h^{-1}\,\mathrm{Mpc}$, and each row corresponding to number densities $\bar{n}=10^{-3}$ and $3\times10^{-4} \, h^{3}\,\mathrm{Mpc}^{-3}$. 
    In the top panels, the theory fits are in solid lines and simulation data points in circular markers. The bottom panels show the residuals of the fits using the standard deviation of simulations as errors and we add gray bands to demonstrate the $1\sigma$ range.
    For all panels, blue, orange, and red colors represent $P_0$, $P_2$, and $M_0$, respectively, and the fits adopt $k_\mathrm{max}^P=0.2\,h\,\mathrm{Mpc}^{-1}$ and $k_\mathrm{max}^M=0.12\,h\,\mathrm{Mpc}^{-1}$. 
    The successful $<1\sigma$ fit for all cases demonstrate that the theoretical precision is enough to be adopted for near-term datasets. Furthermore the fits at different number densities indicate that potential stochastic effects are degenerate with other parameters and can be neglected (\S\ref{sec:stochasticity}). 
    }
    \label{fig:fit_grid}
\end{figure}

In Fig.~\ref{fig:fit_grid} we show the fit against $P_0$, $P_2$, and $M_0$ of periodic boxes, with $k_\mathrm{max}^P=0.2\,h\,\mathrm{Mpc}^{-1}$ and  $k_\mathrm{max}^M=0.12\,h\,\mathrm{Mpc}^{-1}$. Through random downsampling, we fit the catalog at two different number densities $\bar{n}=10^{-3}$ and $3\times10^{-4} \, h^{-3}\,\mathrm{Mpc}^3$, with three different smoothing radii $R=10$, 15, 20 $h^{-1}\,\mathrm{Mpc}$. We will employ the \texttt{scipy} minimizer module in \texttt{Cobaya} \cite{Cobaya,CobayaCode,NelderMead} for the fit. Based on the errors calculated from the 25 boxes, we are able to fit all boxes to $\sim1\sigma$ with the same non-stochastic nuisance parameters (biases $b_i$ and counterterms $\alpha_{2n}$). The volume of each box ($8\,h^{-3}\,\mathrm{Gpc}^3$) is larger than the effective volume of all DESI DR2 redshift bins \cite{DESI-DR2}, indicating that the model accuracy is sufficient for analysis on upcoming data releases, although systematics modeling, such as fiber assignment effects, still must be addressed before data application. 

\subsection{Cutsky mocks}
\label{sec:cutsky}

We use the public, DESI DR1 cutsky mocks \cite{DESI24-II} for validation of the MPS with survey geometry, focusing on the LRG2 redshift bin, spanning $0.6<z<0.8$. The LRG2 mocks are generated based on 25 periodic \textsc{AbacusSummit} simulations \cite{Maksimova21,Garrison18,Garrison21} in cubic boxes of $2\,h^{-1}\, \mathrm{Gpc}$.  The mocks are based on the $z=0.8$ output with galaxies included using the extended HOD model that incorporates velocity bias through two parameters: $\alpha_c$ and $\alpha_s$ \cite{Guo15}.  The first adds a velocity dispersion for central galaxies of $\alpha_c$ times the halo velocity dispersion.  The second scales the satellite-halo relative velocity by $\alpha_s$.  The standard five-parameter HOD model (Eqn.~\ref{eqn:hod-cen} and \ref{eqn:hod-sat}) \cite{Zheng07} can be recovered in the limit $\alpha_c=0$ and $\alpha_s=1$.  The HOD parameters used are \{$\log M_{\mathrm{cut}}=12.64$, $\log M_{1}=13.71$, $\sigma=0.09$, $\alpha=1.18$, $\kappa=0.6$, $\alpha_{c}=0.19$, $\alpha_{s}=0.95$\} which are best-fits to the 3D correlation function and number density in the DESI One-Percent Survey \cite{Yuan23}, modulo the incompleteness fraction. 
The periodic box results are then matched to the DESI NGC and SGC footprints of DR1 by applying coordinate transforms and including the line-of-sight velocities for RSD. This transforms the mocks into a total footprint of $\sim 5800\deg^2$ with the window matrix as shown in Fig.~\ref{fig:window}. 
For this work, we will focus on the mocks without fiber assignment (``complete'' mocks) to capture the survey geometry effects independent of other observational artifacts. The effects of fiber assignment will be investigated in the future when applying the methodology to observational data. 
We refer the reader to ref.~\cite{DESI24-II} for further details about these simulations.

\begin{figure}
    \centering
    \includegraphics[width=0.98\linewidth]{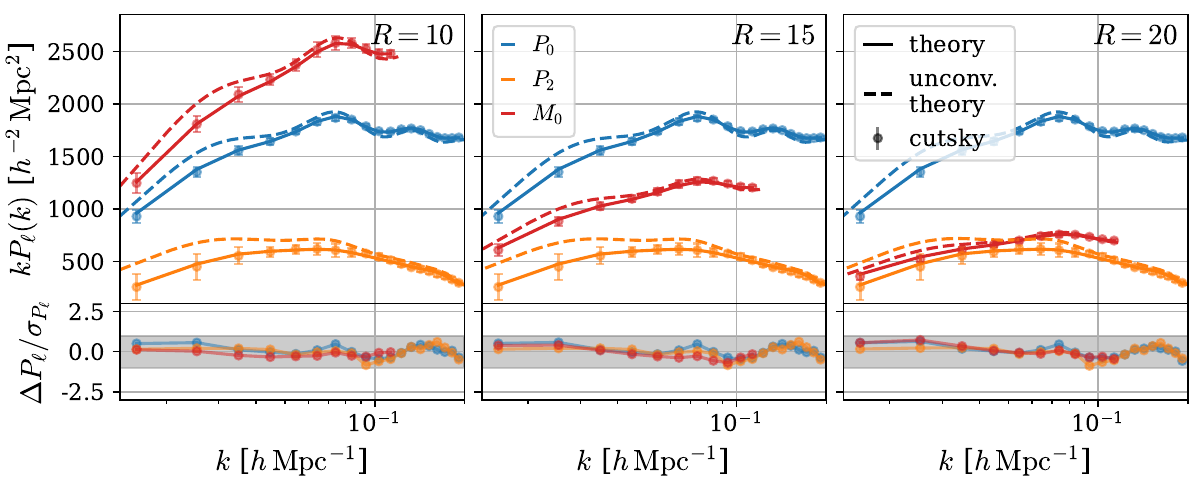}
    \caption{The $P$-$M$ joint fits to DESI DR1 cutsky mocks in the LRG2 ($0.6<z<0.8$) bin with varying smoothing scales $R=10, 15,20\, h^{-1}\,\mathrm{Mpc}$. The top panels show the fits directly, with the window-convolved and unconvolved theory shown in solid and dashed lines against the mock measurements in circular markers. The bottom panels show the residual of the fits using simulation errors, with the $1\sigma$ range highlighted by the gray band. The fits are $<1\sigma$ for all cases, demonstrating that the survey geometry is propagated to the spectrum at precision satisfactory for any near-future data. 
    }
    \label{fig:cutsky_fit}
\end{figure}

We measure both the power spectrum multipoles and MPS monopole from each of the 25 cutsky mocks with smoothing radii of $R=10$, 15, and $20\,h^{-1}\mathrm{Mpc}$, and measure their standard deviation. Using these errors we once again use the \texttt{scipy} minimizer module in \texttt{Cobaya} \cite{Cobaya,CobayaCode,NelderMead} to inform our joint fits of $P$ and $M$. The fit results are shown in Fig.~\ref{fig:cutsky_fit}, showing that one can model the MPS with a realistic survey window jointly with the power spectrum to within $1\,\sigma$ of observational error bars. Note that since these are in DR1 mocks with a smaller and less homogeneous footprint than future data releases, this agreement indicates that we can model such effects for future DESI results.

\section{Alcock-Paczynski Effect}
\label{sec:AP}

In practice, galaxy surveys detect the 3D position of LSS using angles and redshift, which require translation to positions using a fiducial cosmology. As it is unlikely that the fiducial cosmology is identical to the ``true'' cosmology, the choice of cosmology results in a coordinate distortion and one must account for the cosmology mis-specification. This is referred to as the Alcock-Paczynski (A-P) effect \cite{AlcockPaczynski}.

Using the angle on the sky $\Delta \theta$ and redshift difference $\Delta z$, one can compute their perpendicular and los separations using 
\begin{align}
    r_\perp \approx D_A(z) \Delta \theta \qquad , \qquad r_\parallel = \Delta \chi\approx \frac{c \Delta z}{H(z)}
\end{align}
where the comoving distance $\chi$ and angular diameter distance $D_A$ are 
\begin{align}
    \chi = \int \frac{c\,dz'}{H(z')} \qquad , \qquad D_A(z) = \frac{\chi(z)}{(1+z)}
\end{align}
This implies that the fiducial and true wavevectors are related by
\begin{align}
    \mathbf{k}_\mathrm{fid} = \mathbf{k}_\parallel \alpha_\parallel(z) + \mathbf{k}_\perp \alpha_\perp(z)
    \label{eqn:kAP}
\end{align}
for $\alpha_\parallel=H(z)_\mathrm{fid}/H(z)_\mathrm{true}$ and $\alpha_\perp=D_A(z)_\mathrm{true}/D_A(z)_\mathrm{fid}$. 
Accounting for the units of volume in the power spectrum ($P_\mathrm{obs}(\mathbf{k}_\mathrm{fid},z) d^3 k_\mathrm{fid}=P_\mathrm{true}(\mathbf{k}_\mathrm{true},z) d^3 k_\mathrm{true}$) we get \cite{AlcockPaczynski,Padmanabhan08} 
\begin{equation}
    P_\mathrm{obs}(\mathbf{k}_\mathrm{fid},z) = \alpha_\parallel^{-1}(z)\alpha_\perp^{-2}(z) P_\mathrm{true}(\mathbf{k}_\mathrm{true},z)
\end{equation}
The inclusion of A-P effects in the power spectrum has been standardized, such that the effect can be calculated at every cosmology analytically (to lowest order in $\delta\alpha$) without significant loss in the MCMC inference pipeline.

For the MPS the fiducial cosmology dependence is more complex than for the power spectrum, due to both the involvement of an additional overdensity field and the smoothing $W_R(k)$.
An analytical solution can be best captured by recalling the expression of $M$ as an integral over $B$ (Eq.~\ref{eqn:bispectrum-integral}).
The cosmology dependence of the (tree-level) bispectrum is routinely included and can be calculated extremely rapidly; for example with software such as \texttt{FOLPS-D}\footnote{\url{https://github.com/alejandroaviles/folpsD}} \cite{Noriega22} the evaluation of $B$ takes less than a millisecond. In practice, however, incorporating this effect without loss in computational time is non-trivial, as the integration over $B$ is the computationally expensive procedure in evaluating $M$.  While some of this cost can be handled through the use of emulators, we also wanted to explore an alternative route for incorporating the A-P effect.  Below we shall demonstrate that the change of MPS over the fiducial cosmology choice is smooth enough that we can interpolate between cosmologies to match the observed cosmology to the evaluated cosmology at every step of the MCMC chain \cite{White15}.  This way there will be no position distortion, as the cosmologies used for MPS measurement and model evaluation are identical (alternatively we can use the inverse scaling to introduce the distortion into the theory). Parameter estimation runtime is not compromised significantly, as the measurement pipeline is run once over a sufficiently large cosmology parameter space and then the results interpolated during the MCMC steps.

The MPS shares the same units of volume as the power spectrum, so the volume factor $\alpha_\parallel^{-1}\alpha_\perp^{-2}$ can be factored out in an identical manner. We therefore focus exclusively on the residual transformation arising from the cosmological dependence of $\mathbf{k}$.
In a $\Lambda$CDM cosmology, the only cosmological parameter that A-P depends on is $\Omega_m$. For extended cosmology models, there can be more parameter dependencies. For instance, in a $w_0w_a$CDM, which has been found to be preferred by DESI+CMB constraints \cite{DESI-DR2}, there are additional dependencies on the dark energy equation of state $w_0$, $w_a$.  These also induce very smooth distortions.

\begin{figure}
    \centering
    \includegraphics[width=0.98\linewidth]{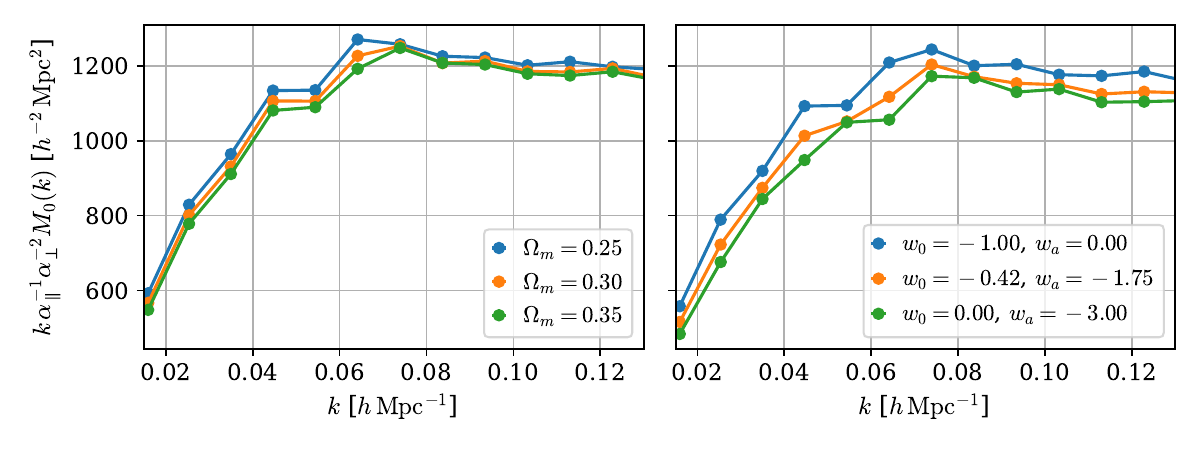}
    \caption{The residual A-P effect of $M$ in $\Lambda$CDM (left) and $w_0 w_a$CDM (right), using a smoothing radius $R=15\,h^{-1}\,\mathrm{Mpc}$ and one instance of DESI DR1 cutsky mocks.  For $\Lambda$CDM $\Omega_m$ is varied between $0.25$ and $0.35$. The range for $w_0w_a$CDM correspond to the $-2\sigma$ to $+2\sigma$ extent in the recent DESI+CMB constraints \cite{DESI-DR2}. Over these cosmologies the change in $\alpha_\parallel^{-1}\alpha_\perp^{-2}\,M_0$ is small and smooth.}
    \label{fig:AP1}
\end{figure}

Figure \ref{fig:AP1} shows the residual A-P variation in both $\Lambda$CDM and $w_0w_a$CDM, varying $\Omega_m$ between $0.25$ and $0.35$ and $w_0w_a$ within recent $-2\sigma$ to $+2\sigma$ constraints with DESI+CMB \cite{DESI-DR2}. The variation of $\alpha_\parallel^{-1}\alpha_\perp^{-2}\,M_0$ over these wide range of cosmologies is small and smooth, implying that it can be interpolated easily. 

\section{Conclusions}
\label{sec:conclusion}

Higher-order statistics of large-scale structure offer insight into non-Gaussian information that is not captured in traditional two-point correlators. As cosmological surveys make advance in precision, accessing these non-Gaussian information becomes important to break parameter degeneracies and improve cosmological constraints \cite{Chudaykin25a,Chudaykin25b}. The marked power spectrum (MPS) \cite{White16,Philcox21,Ebina24} probes particular compressions of higher-order correlations while retaining the structure of two-point correlators, allowing much of the existing two-point analysis infrastructure to be reused and enabling perturbative modeling without introducing new theoretical uncertainties. In this work we address both the perturbative and survey modeling of the MPS, in order to prepare for application in near-future datasets. 

Based on the analytical model, we redefine the MPS in order to isolate the higher-order information and decrease overlap with the power spectrum. Through this redefinition we find strong potential for degeneracy breaking, with reasonably expected changes in the secondary bias $b_2$ yielding MPS that are different in amplitude by a factor of 2, which is in agreement with, but strengthens, the results of ref.~\cite{Ebina24}. This signal exceeds observational errors and higher-order corrections by more than an order of magnitude. 

Recent work raised the possibility that the MPS may have additional stochasticity from constructing the density field from a finite number of objects \cite{Karcher24}. We show this source of noise is already encompassed in the model's free parameters and use 25 mock catalogs in $2\,h^{-1}\mathrm{Gpc}$ periodic boxes to show that this is not a concern. We also demonstrate that small higher-loop corrections to the MPS are not needed for the fits in this work, though they may be required in future analyses \cite{Philcox21,Ebina24}.

The effect of survey geometry on the MPS can be modeled similarly to the power spectrum, since the mark is a local, smooth function. 
This is a major advantage in terms of data application, as this simplifies the modeling and infrastructure to implement these effects already exists and is well-known in two-point analyses \cite{DESI-DR2,DESI24-V}.  We show how to modify standard 2-point function codes to compute the MPS and its window matrix, demonstrating that converged results can be obtained with modest computational requirements. 
We test the modeling of the survey geometry by jointly fitting the power spectrum and MPS against DESI DR1 cutsky catalogs. The results show consistency between model and simulation within $1\sigma$ in all cases, up to $k^P_\mathrm{max}=0.2\,h\,\mathrm{Mpc}^{-1}$ for the power spectrum multipoles and $k^M_\mathrm{max}=0.12\,h\,\mathrm{Mpc}^{-1}$ for the MPS monopole. As the DESI DR1 data are more non-uniform than future DESI data releases, we expect that this agreement will hold for all near-future datasets. 

Another major advantage of the MPS is the (relatively) small number of data points, as it has one free momentum vector constrained to low $k$. This will especially help in the construction of covariance matrices, as the number of simulations necessary to construct the covariance is likely similar to that of the power spectrum. While a full investigation requires a large simulation suite and is left for future work, we develop insight through several avenues. Our redefinition of the MPS decreases the covariance between $P$ and $M$, lowering the required precision of their cross-covariance. Direct calculation of the covariance reveals that the Gaussian approximation captures all diagonal contributions, in agreement with ref.~\cite{Harscouet24}, but that the cross-covariance is not strongly suppressed, unlike the bispectrum case \cite{Biagetti22,Salvalaggio24} due  to difference in mode-counting factors. We qualitatively confirm these findings with our 25 periodic box simulations.

We also address fiducial cosmology dependence. While evaluating the Alcock-Paczynski effect \cite{AlcockPaczynski} for the MPS is technically possible, implementing it without significant computational cost is non-trivial due to the additional density field and smoothing kernel $W_R(k)$. This can be circumvented by re-evaluating the redshift-distance relation at every likelihood stage. We demonstrate that the cosmology dependence of the MPS is marginal and smooth, allowing simple interpolation over cosmologies to perform this at high precision.

Further work is still required before application to data. The effects of fiber assignment must be modeled, as redshift surveys like DESI do not obtain redshifts for all galaxies and fiber completeness varies with environmental variables such as local density. We must also validate that the MPS returns unbiased cosmological constraints from mock catalogs, which requires faster analytical techniques such as FFTlog \cite{Ham00} or cosmology-based emulators \cite{DESI24-V}, or both. The covariance question must likewise be addressed with a large simulation suite providing precision necessary for modern datasets.

\section{Data Availability}

The software used for the analytical calculations in this work are publicly available at \href{https://github.com/HarukiEbina/markedPS}{\texttt{https://github.com/HarukiEbina/markedPS}}. 

\section*{Acknowledgements}

HE and MW were supported by the DOE.  This research was supported in part by grant NSF PHY-2309135 to the Kavli Institute for Theoretical Physics (KITP).  This work made use of the Cobaya analysis code \cite{Cobaya,CobayaCode}.
This research used resources of the National Energy Research Scientific Computing Center (NERSC), a Department of Energy User Facility.

\appendix
\section{The beyond-2pt information in $\mathcal{M}$}
\label{sec:b2pinfo}

As described in Eqn.~\ref{defMorig}, the original marked power spectrum $\mathcal{M}$ is a combination of two-point and beyond two-point information. For brevity, we will summarize the two-point information as $\mathcal{M}_\mathrm{2pt}$
\begin{align}
    \mathcal{M} = \mathcal{M}_\mathrm{2pt} + M
\end{align}
where $M$ is the marked power spectrum in the main text, extracting the beyond two-point information explicitly.
Here, we will explore the respective fraction of these two components in order to gain better perspective of statements made in past work referencing $\mathcal{M}$ \cite{White16,Philcox21,Ebina24}. To do this, we measure $\mathcal{M}_\mathrm{2pt}$ and $M$ for a high-density ($\bar{n}=10^{-3}\,h^3\,\mathrm{Mpc}^{-3}$) DESI LRG-like mock catalogs over 25 periodic simulation boxes of volume $8\,h^{-3}\,\mathrm{Gpc}^3$ at $z=0.8$ (see \S\ref{sec:periodicbox}).

\begin{figure}
    \centering
    \includegraphics[width=0.98\linewidth]{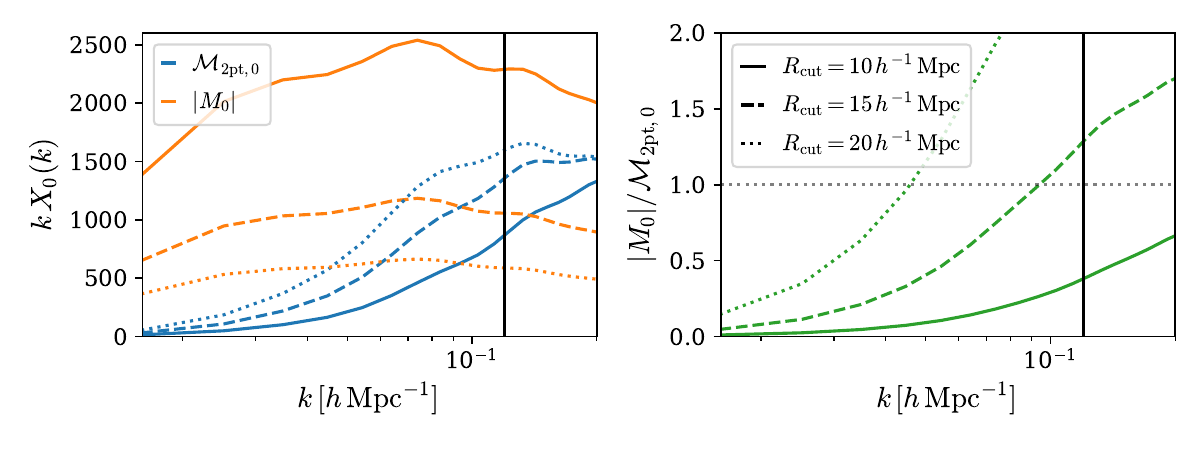}
    \caption{The two-point ($\mathcal{M}_\mathrm{2pt}$) and beyond-two-point ($M$) power in $\mathcal{M}_0$. The left panel directly shows the $\mathcal{M}_{\mathrm{2pt},0}$ (blue) and $|M_0|$ (orange) as measured from mock catalogs in periodic boxes, with the smoothing scales $R=10$ (solid), 15 (dashed) and 20$h^{-1}\,\text{Mpc}$ (dotted). The right panel shows the ratio of the two components $|M_0|/\mathcal{M}_{\mathrm{2pt},0}$ (green), with the same line styles corresponding to smoothing radii as the left panel. As anticipated the choice of mark suppresses the two-point component at low-$k$, with the rate of increase in two-point power strongly dependent on the smoothing radius.}
    \label{fig:M_2pt_3pt_comparison}
\end{figure}

Of course, as the MPS does not break the perturbative scaling, for an unspecified mark the linear term in $\mathcal{M}_\mathrm{2pt}$ will dominate. However, taking advantage of the overall scaling of $\mathcal{M}_\mathrm{2pt}$ \cite{Philcox21,Ebina24}
\begin{align}
    \mathcal{M}_{\mathrm{2pt},\ell}(k) &= C_{\delta_M}^a(k) C_{\delta_M}^b(k) P^{\rm one-loop}_\ell(k) \\
    \mathrm{for} &\quad C_{\delta_M}^a(k) = C_0^a + C_1^a W_R(k)
\end{align}
one can set one or both of the marks to $m=1-\delta_{g,R}$ to suppress two-point information at theoretically well-described large-scales (low-$k$).  This was the approach adopted in ref.~\cite{Ebina24}. Using the cross-correlation of this mark with the unmarked density field for $R=10$, 15, and 20 $h^{-1}\,\mathrm{Mpc}$\footnote{As mentioned in \S\ref{sec:mps}, this yields the same $M$ as using $m=1+\delta_{g,R}$, as done in the main text}, we find the results in in Fig.~\ref{fig:M_2pt_3pt_comparison}. 
As intended, the mark choice suppresses the $\mathcal{M}_\mathrm{2pt}$ sufficiently such that $M$ dominates at low-$k$. This decomposition, however, also shows that the two-point information provides a non-negligible contribution to even linear scales, which can introduce significant covariance between $P$ and $\mathcal{M}$ and in turn make the requirements on the covariance more stringent. Thus, it is better motivated to directly access the beyond two-point term $M$ by subtracting $\mathcal{M}_{\rm 2pt}$ at the level of the estimator.

\section{Choice of marks}
\label{sec:stability}

\begin{figure}
    \centering
    \includegraphics[width=0.98\linewidth]{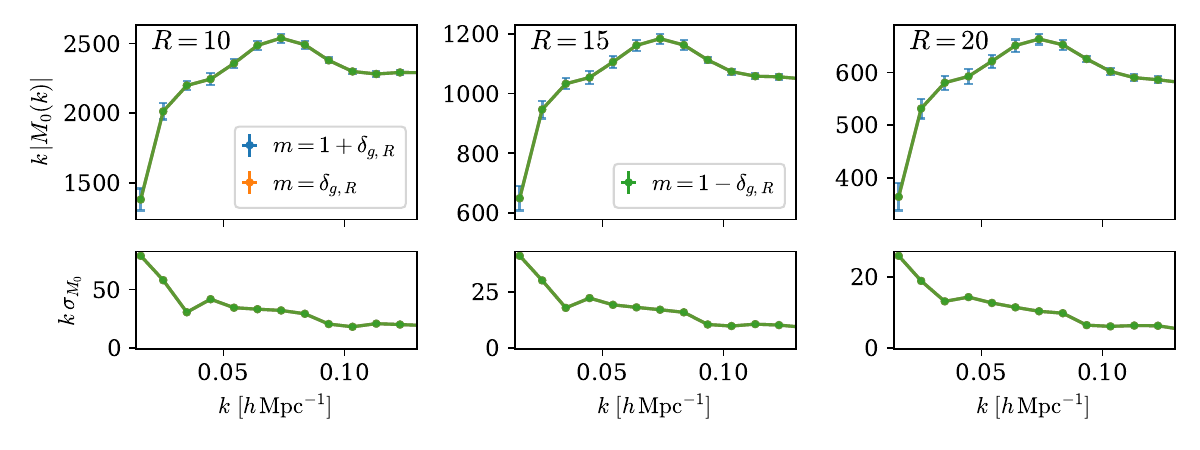}
    \caption{$|M_0|$ and its standard deviation $\sigma_{M_0}$ for the cross-spectrum of $m=1\pm\delta_{g,R}$ and $m=\delta_{g,R}$ for $R=10$, 15, and $20h^{-1}\text{Mpc}$, measured over 25 $8\,h^{-3}\,\mathrm{Gpc}^3$ boxes. The three marks give identical means (up to a sign) and dispersion as expected from Eqn.~\ref{eqn:defM}.  In the figure the blue, orange and green points and lines lie on top of each other.
    }
    \label{fig:stability}
\end{figure}

The calculation in Eqn.~\ref{eqn:defM} indicates that the cross-spectrum of the unmarked field with either $m=1\pm\delta_{g,R}$ or $m=\delta_{g,R}$ agree, up to an overall sign. 
Ref.~\cite{Ebina24} found that there are practical differences, with the spectra of $m=1+\delta_{g,R}$ more stable than that of $m=1-\delta_{g,R}$, especially for small smoothing scales (e.g.~$R=10h^{-1}\text{Mpc}$) where the long tail of $\delta_{g,R}$ can lead to zero-crossings of the mark. By contrast, we find that the cross-spectrum of $m=1\pm\delta_{g,R}$ and $m=\delta_{g,R}$ are in agreement with each other, as shown in Fig.~\ref{fig:stability}. This may be a result of a more stable observation code, with the density estimation following that of the reconstruction code \texttt{pyrecon}. 
The situation may be different for scenarios involving systematics and systematics correction weights, such as fiber assignment effects. The correction schemes have been tested for subsets of the data, but under significant reweighting it is unclear whether the corrections continue to work as anticipated. In these scenarios, it is conceivable that additional constants in the mark can help counter systematics due to less relative reweighting. 
Furthermore, one can optimize the mark in a higher polynomial order to tighten specific constraints \cite{Cowell24}. We will leave such investigation to future work.

\section{Scale-dependence of $M_{13}$}
\label{sec:Mshape}

Here we discuss the similarity between the scale-dependence of $M_{13}$ and $P_L$, as seen in Fig.~\ref{fig:degeneracy}. 
We have discussed in \S\ref{sec:mps} that the $M_{13}$ contribution consists of $\expval{\delta_{g,R,1}^{(1)}\delta_{g,1}^{(2)}\delta_{g,2}^{(1)}}$ and $\expval{\delta_{g,R,1}^{(2)}\delta_{g,1}^{(1)}\delta_{g,2}^{(1)}}$ that merely differ by the argument of the smoothing kernel. 
Focusing on the first term, we have
\begin{equation}
    \expval{\delta_{g,R,1}^{(1)}\delta_{g,1}^{(2)}\delta_{g,2}^{(1)}} = \int_\mathbf{p} Z_2(\mathbf{k},\mathbf{p}) Z_1(\mathbf{k}) Z_1(\mathbf{p}) W_R(p)P_L(k)P_L(p) 
\end{equation}
in the absence of stochastic terms. In \S\ref{sec:mps} and Fig.~\ref{fig:degeneracy} we have already demonstrated that the $b_2$ contributions are approximately $\propto P_L(k)$, with little additional shape ($k$) dependence, and the $b_s$ contribution is subdominant to that of $b_2$ by a factor of 5 or more. Now we inspect the other terms. As shown in Eqns.~\ref{eqn:F2} and \ref{eqn:G2}, the first two terms of the kernels $F_2$ and $G_2$ merely contribute angular dependencies that will only matter when considering the marked spectrum quadrupole. The final terms are the only contributors to spectrum shape for both kernels. For the $F_2$ contribution, we find, however, that the final term vanishes due to symmetry, i.e.
\begin{align}
    \expval{\delta_{g,R,1}^{(1)}\delta_{g,1}^{(2)}\delta_{g,2}^{(1)}} &\supset \int_\mathbf{p} b_1 F_2(\mathbf{k},\mathbf{p}) Z_1(\mathbf{k}) Z_1(\mathbf{p}) W_R(p)P_L(k)P_L(p) \\
    &= b_1 Z_1(\mathbf{k}) P_L(k) \int W_R(p) P_L(p) p^2 dp \\ 
    &\quad\times \bigg[\frac{5}{7} \left(\frac{4\pi}{3}(3b_1+f)\right) + \frac{2}{7}\left(\frac{4\pi}{15}(5b_1 + f + 2f\mu^2)\right) + \frac{1}{2}\times 0\bigg]
\end{align}
adding no additional scale-dependence. The $G_2$ contribution offers a small correction to the shape, with the monopole contribution being 
\begin{align}
    \expval{\delta_{g,R,1}^{(1)}\delta_{g,1}^{(2)}\delta_{g,2}^{(1)}}_0 &\supset \int_\mathbf{p} f\mu_{\mathbf{k}+\mathbf{p}}^2 G_2(\mathbf{k},\mathbf{p}) Z_1(\mathbf{k}) Z_1(\mathbf{p}) W_R(p)P_L(k)P_L(p) \\
    &= P_L(k) \int \frac{p^2\, dp}{(2\pi)^2} \left(c_0 + c_2 \frac{k^2}{p^2} + c_4 \frac{k^4}{p^4} + \mathcal{O}(k^6/p^6) \right) W_R(p) P_L(p) 
\end{align}
with
\begin{align}
    c_0 &= \frac{2f}{11025} \left(2275 b_1^2 + 2394 b_1 f+ 603 f^2\right) \approx 2.269 \\
    c_2 &= \frac{-128f^2}{25725} \left(7 b_1 + 3f \right) \approx -0.067 \\
    c_4 &= \frac{256 f^2}{77125} \left(7 b_1 + 3f\right)
    \approx 0.045
\end{align}
where the numerical approximations are for $b_1=2$ and $f=0.9$. Finally the last contribution to $Z_2$ is
\begin{align}
    \expval{\delta_{g,R,1}^{(1)}\delta_{g,1}^{(2)}\delta_{g,2}^{(1)}} &\supset \int_\mathbf{p} \frac{fk\mu_{\mathbf{k}+\mathbf{p}}}{2}\left[\frac{\mu}{k}(b_1+f\mu_p^2) + \frac{\mu_p}{k_p}(b_1 + f\mu^2) \right] Z_1(\mathbf{k}) Z_1(\mathbf{p}) W_R(p)P_L(k)P_L(p) \\
    = &Z_1(\mathbf{k})P_L(k) \left( \frac{b_1 f}{15}(5b_1+3f) + \left(b_1^2 f+ b_1 f^2 +\frac{2f^3}{5} \right) \mu^2 \right) \int\frac{p^2 \, dp}{(2\pi)^2}  W_R(p)P_L(p)
\end{align}
again indicating that the scale-dependence of $M$ resembles that of $P_L$.

Although the triangle configurations are different, since $k\lesssim0.08\,h\,\mathrm{Mpc}^{-1}$ and $p\lesssim1/R<0.1\,h\,\mathrm{Mpc}^{-1}$, the situation here is mathematically similar to the squeezed bispectrum, which also display similar scale-dependence to the power spectrum. For the squeezed bispectrum, one operates with one of the triangle sides $q$ `squeezed' to zero in momentum space
\begin{equation}
    \lim_{q\to 0} B(\mathbf{k},-\mathbf{k},\mathbf{q}) = Z_1^2(\mathbf{k}) Z_2(\mathbf{k},-\mathbf{k}) P_L^2(k) + Z_1(\mathbf{k}) P_L(k) Z_1(\mathbf{q}) P_L(q) \left[Z_2(\mathbf{k},\mathbf{q})+ Z_2(-\mathbf{k},\mathbf{q})\right]
\end{equation}
where we use $Z_1(\mathbf{k}) =Z_1(-\mathbf{k})$ to simplify the expression.  Since $Z_2(\mathbf{k},-\mathbf{k})=b_2/2$, the first term is simply $(b_2/2) \, Z_1^2(\mathbf{k}) P_L^2(k)$. The sum of $Z_2$'s in the second term cancels any term that is odd\footnote{Even if $q>0$, all terms odd in $\hat{k}$ will vanish under angular integration due to symmetry.} in $\widehat{k}$, i.e.
\begin{equation}
    Z_2(\mathbf{k},\mathbf{q})+Z_2(-\mathbf{k},\mathbf{q}) = 2\left[\frac{5}{7} + \frac{2}{7}\left(\widehat{k}\cdot\widehat{q}\right)^2 + \mu^2 \left(\frac{3}{7} + \frac{4}{7}\left(\widehat{k}\cdot\widehat{q}\right)^2\right) + \frac{b_2}{2} + b_s \left( \left(\widehat{k}\cdot\widehat{q}\right)^2-\frac{1}{3}\right)\right]
\end{equation}
This leaves no $k$ dependence to alter the spectrum shape from $Z_1(\mathbf{k})P_L(k)Z_1(\mathbf{q})P_L(q)$.
This result is often quoted as the squeezed matter bispectrum $\lim_{q\to0} B_m(k,k,q)\propto P_L(k)P_L(q)$, as the first term vanishes in the absence of $b_2$. 
Note that this computation does not consider the stochastic and counterterm contributions to the the spectra, which will de-correlate the shapes of MPS and squeezed bispectrum from $P_L$. 

\section{Covariance matrix}
\label{app:covariance}

Here, we describe the calculation for the variance of $M$ and covariance between $P$ and $M$, supplementing the material in \S\ref{sec:covariance}. 

\subsection{Covariance of $M$}
\label{app:variance}

As introduced in the main text, the $k-k^\prime$ covariance of a marked spectrum can be calculated as 
\begin{equation}
    \mathrm{Cov}\left(\widehat{M}_k, \widehat{M}_{k'}\right) \equiv \expval{\widehat{M}_k \widehat{M}_{k'}} - \expval{\widehat{M}_k}\expval{ \widehat{M}_{k'}}
\end{equation}
Expanding this expression we obtain
\begin{align}
    \mathrm{Cov}\left(\widehat{M}_k, \widehat{M}_{k'}\right) &= 
    V^{-2} \int_{V_k} \frac{d^3q}{V_k}  \int_{V_{k'}} \frac{d^3q'}{V_{k'}} \int_\mathbf{p} \int_{\mathbf{p}'} W_R(p) W_R(p') \\
    &\qquad (\left\langle \delta_{-\mathbf{q}} \delta_{\mathbf{p}} \delta_{\mathbf{q}-\mathbf{p}} \delta_{-\mathbf{q}'} \delta_{\mathbf{p}'} \delta_{\mathbf{q}'-\mathbf{p}'}\right\rangle 
    - \left\langle\delta_{-\mathbf{q}} \delta_{\mathbf{p}} \delta_{\mathbf{q}-\mathbf{p}}\right\rangle 
    \left\langle\delta_{-\mathbf{q}'} \delta_{\mathbf{p}'} \delta_{\mathbf{q}'-\mathbf{p}'} \right\rangle) \\
    &= \mathcal{I}[W_R(p) W_R(p')\left\langle \delta_{-\mathbf{q}} \delta_{\mathbf{p}} \delta_{\mathbf{q}-\mathbf{p}} \delta_{-\mathbf{q}'} \delta_{\mathbf{p}'} \delta_{\mathbf{q}'-\mathbf{p}'}\right\rangle ] - M_k M_{k'}
\end{align}
where we define\footnote{This differs from that in ref.~\cite{Harscouet24} by factors of $W_R$, due to the aforementioned loss of symmetry between fields in $\tilde\delta_M$.}
\begin{equation}
    \mathcal{I}[f] = V^{-2}\int_{V_k} \frac{d^3q}{V_k}\int_{V_{k'}} \frac{d^3q'}{V_{k'}} \int_\mathbf{p} \int_{\mathbf{p}'} f(\mathbf{q},\mathbf{q}',\mathbf{p},\mathbf{p}')
\end{equation}

As discussed in the main text, there are three possible categories of contractions: $(2,2,2)$ (e.g.~$\expval{ab}\expval{cd}\expval{ef}$), $(3,3)$ (e.g.~$\expval{abc}\expval{def}$), $(4,2)$ (e.g.~$\expval{abcd}\expval{ef}$), and $(6)$ (e.g.~$\expval{abcdef}$), for fields $a=\delta_{-\mathbf{q}}$, $b=\delta_{\mathbf{p}}$, $c=\delta_{\mathbf{q}-\mathbf{p}}$, $d=\delta_{-\mathbf{q}'}$, $e=\delta_{\mathbf{p}'}$, $f=\delta_{\mathbf{q}'-\mathbf{p}'}$. We will enumerate all possible contractions for each category.

Before we start, it will be useful to point out that $b=\delta_{\mathbf{p}}$ and $c=\delta_{\mathbf{k}-\mathbf{p}}$ are equivalent by the transformation $\mathbf{p}\leftrightarrow \mathbf{q}-\mathbf{p}$, which will help simplify our calculations. Note that the full expression does not adhere to this symmetry, as $W(p)$ will transform to $W(|\mathbf{q}-\mathbf{p}|)$. A similar symmetry is present for $e=\delta_{\mathbf{p}'}$ and $f=\delta_{\mathbf{q}'-\mathbf{p}'}$, with the transformation $\mathbf{p}'\leftrightarrow \mathbf{q}'-\mathbf{p}'$.

For the $(2,2,2)$ contractions there are 15 terms
\begin{align}
\langle abcdef \rangle_{222} = 
\langle ab \rangle \langle cd \rangle \langle ef \rangle &+ \langle ab \rangle \langle ce \rangle \langle df \rangle + \langle ab \rangle \langle cf \rangle \langle de \rangle \, + \nonumber\\
\langle ac \rangle \langle bd \rangle \langle ef \rangle &+ \langle ac \rangle \langle be \rangle \langle df \rangle + \langle ac \rangle \langle bf \rangle \langle de \rangle \, + \nonumber\\
\langle ad \rangle \langle bc \rangle \langle ef \rangle &+ \langle ad \rangle \langle be \rangle \langle cf \rangle + \langle ad \rangle \langle bf \rangle \langle ce \rangle \, + \nonumber\\
\langle ae \rangle \langle cd \rangle \langle bf \rangle &+ \langle ae \rangle \langle bc \rangle \langle df \rangle + \langle ae \rangle \langle cf \rangle \langle bd \rangle \, + \nonumber\\
\langle af \rangle \langle cd \rangle \langle eb \rangle &+ \langle af \rangle \langle ce \rangle \langle db \rangle + \langle af \rangle \langle bc \rangle \langle de \rangle
\end{align}
Notice that the three fields involved in constructing each $M$ ($abc$ and $def$) necessarily have zero net momentum. As correlators enforce that the sum of field momenta are zero, if two of these fields are contracted (e.g.~$\expval{ab}$) the momentum of the third field (e.g.~$c$) must be zero, and thus contributes no power. This reduces the number of terms to 6
\begin{align}
\langle abcdef \rangle_{222} = 
\langle ad \rangle \langle be \rangle \langle cf \rangle &+ \langle ad \rangle \langle bf \rangle \langle ce \rangle \, + \langle ae \rangle \langle cd \rangle \langle bf \rangle + \nonumber\\
\langle ae \rangle \langle cf \rangle \langle bd \rangle &+ 
\langle af \rangle \langle cd \rangle \langle eb \rangle + \langle af \rangle \langle ce \rangle \langle db \rangle 
\end{align}

The first two terms that contract the two external momenta ($\expval{ab}=\left\langle \delta_{-\mathbf{q}} \delta_{-\mathbf{q}'} \right\rangle$) contribute diagonal terms of the covariance. Let us start with the first term
\begin{align}
    \langle ad \rangle \langle be \rangle \langle cf \rangle 
    &= \mathcal{I}\left[ W_R(p)W_R(p')
    \left\langle \delta_{-\mathbf{k}} \delta_{-\mathbf{k}'} \right\rangle
    \left\langle \delta_{\mathbf{p}}\delta_{\mathbf{p}'} \right \rangle
    \left\langle \delta_{\mathbf{k}-\mathbf{p}} \delta_{\mathbf{k}'-\mathbf{p}'}\right\rangle \right] \\
    &= (2\pi)^9 V^{-2} \int_{V_k}\frac{d^3q}{V_k}\int_{V_{k'}}\frac{d^3q'}{V_{k'}} 
    \int_{\mathbf{p}}\int_{\mathbf{p}'} W_R(p) W_R(p') \nonumber \\
    &\qquad\qquad \delta_D(\mathbf{q}+\mathbf{q}')\delta_D(\mathbf{p}+\mathbf{p}')\delta_D(\mathbf{q}-\mathbf{p}+\mathbf{q}'-\mathbf{p}')
    P(-\mathbf{q})P(\mathbf{p})P(\mathbf{q}-\mathbf{p})
\end{align}
One of the Dirac delta functions yield a $\delta_D(0)=V/(2\pi)^3$, which cancels a volume factor. The other yields $\int_{V_k}d^3k' \delta_D(\mathbf{k}-\mathbf{k}')=\delta^K_{k,k'}$. Using the third delta function to eliminate one integral over an internal momentum and approximating $P(\mathbf{q})\approx P(\mathbf{k})$ we obtain
\begin{align}
    \langle ad \rangle \langle be \rangle \langle cf \rangle 
    &= (2\pi)^3 V^{-1} V_k^{-1}\delta^K_{k,k'} \int_\mathbf{p} W_R^2(p) P(\mathbf{k})P(\mathbf{p})P(\mathbf{k}-\mathbf{p})
\end{align}
The second term has a similar calculation. By taking advantage of the transformation $\mathbf{p}'\leftrightarrow \mathbf{q}'-\mathbf{p}'$, we simplify the expression to be identical to the first term, with the exception of the argument of $W_R$
\begin{align}
    \langle ad \rangle \langle bf \rangle \langle ce \rangle 
    &= \mathcal{I}\left[ W_R(p)W_R(p')
    \left\langle \delta_{-\mathbf{q}} \delta_{-\mathbf{q}'} \right\rangle
    \left\langle \delta_{\mathbf{p}}\delta_{\mathbf{q}'-\mathbf{p}'} \right \rangle
    \left\langle \delta_{\mathbf{q}-\mathbf{p}}\delta_{\mathbf{p}'} \right\rangle \right] \\
    &= \mathcal{I}\left[ W_R(p)W_R(|\mathbf{q}'-\mathbf{p}'|)
    \left\langle \delta_{-\mathbf{q}} \delta_{-\mathbf{q}'} \right\rangle
    \left\langle \delta_{\mathbf{p}}\delta_{\mathbf{p}'} \right \rangle
    \left\langle \delta_{\mathbf{q}-\mathbf{p}} \delta_{\mathbf{q}'-\mathbf{p}'}\right\rangle \right] \\
    &= \langle ad \rangle \langle be \rangle \langle cf \rangle  \times \frac{W_R(|\mathbf{k}'-\mathbf{p}'|)}{W_R(p')} \nonumber \\
    &= (2\pi)^3 V^{-1} V_k^{-1}\delta^K_{k,k'} \int_\mathbf{p} 
    W_R(|\mathbf{k}-\mathbf{p}|)W_R(p) P(\mathbf{k})P(\mathbf{p})P(\mathbf{k}-\mathbf{p})
\end{align}

The other four terms contribute non-diagonal terms by contracting the external and internal momenta. We can simplify these terms using the transformations $\mathbf{p}\leftrightarrow \mathbf{k}-\mathbf{p}$ and $\mathbf{p}'\leftrightarrow \mathbf{k}'-\mathbf{p}'$ that we have pointed out above. 
\begin{align}
    \langle ae \rangle \langle cd \rangle \langle bf \rangle 
    &= \mathcal{I}\left[ W_R(p)W_R(p')
    \left\langle \delta_{-\mathbf{q}} \delta_{\mathbf{p}'} \right\rangle
    \left\langle \delta_{\mathbf{q}-\mathbf{p}}\delta_{-\mathbf{q}'} \right \rangle
    \left\langle \delta_{\mathbf{p}}\delta_{\mathbf{q}'-\mathbf{p}'} \right\rangle \right] \\
    \langle ae \rangle \langle cf \rangle \langle bd \rangle 
    &= \mathcal{I}\left[ W_R(p)W_R(p')
    \left\langle \delta_{-\mathbf{q}} \delta_{\mathbf{p}'} \right\rangle
    \left\langle \delta_{\mathbf{q}-\mathbf{p}}\delta_{\mathbf{q}'-\mathbf{p}'} \right \rangle
    \left\langle \delta_{\mathbf{p}}\delta_{-\mathbf{q}'} \right\rangle \right] \\
    &= \mathcal{I}\left[ W_R(|\mathbf{q}-\mathbf{p}|)W_R(p')
    \left\langle \delta_{-\mathbf{q}} \delta_{\mathbf{p}'} \right\rangle
    \left\langle \delta_{\mathbf{q}-\mathbf{p}}\delta_{-\mathbf{q}'} \right \rangle
    \left\langle \delta_{\mathbf{p}}\delta_{\mathbf{q}'-\mathbf{p}'} \right\rangle \right] \\
    \langle af \rangle \langle cd \rangle \langle be \rangle  
    &= \mathcal{I}\left[ W_R(p)W_R(p')
    \left\langle \delta_{-\mathbf{q}} \delta_{\mathbf{q}'-\mathbf{p}'} \right\rangle
    \left\langle \delta_{\mathbf{q}-\mathbf{p}}\delta_{-\mathbf{q}'} \right \rangle
    \left\langle \delta_{\mathbf{p}}\delta_{\mathbf{p}'} \right\rangle \right] \\
    &= \mathcal{I}\left[ W_R(p)W_R(|\mathbf{q}'-\mathbf{p}'|)
    \left\langle \delta_{-\mathbf{q}} \delta_{\mathbf{p}'} \right\rangle
    \left\langle \delta_{\mathbf{q}-\mathbf{p}}\delta_{-\mathbf{q}'} \right \rangle
    \left\langle \delta_{\mathbf{p}}\delta_{\mathbf{q}'-\mathbf{p}'} \right\rangle \right] \\
    \langle af \rangle \langle ce \rangle \langle bd \rangle  
    &= \mathcal{I}\left[ W_R(p)W_R(p')
    \left\langle \delta_{-\mathbf{q}} \delta_{\mathbf{q}'-\mathbf{p}'} \right\rangle
    \left\langle \delta_{\mathbf{q}-\mathbf{p}}\delta_{\mathbf{p}'} \right \rangle
    \left\langle \delta_{\mathbf{p}}\delta_{-\mathbf{q}'} \right\rangle \right] \\
    &= \mathcal{I}\left[ W_R(|\mathbf{q}-\mathbf{p}|)W_R(|\mathbf{q}'-\mathbf{p}'|)
    \left\langle \delta_{-\mathbf{q}} \delta_{\mathbf{p}'} \right\rangle
    \left\langle \delta_{\mathbf{q}-\mathbf{p}}\delta_{-\mathbf{q}'} \right \rangle
    \left\langle \delta_{\mathbf{p}}\delta_{\mathbf{q}'-\mathbf{p}'} \right\rangle \right] 
\end{align}
Thus, the sum of the four terms are
\begin{align}
    \tilde N_{222} &=
    \langle ae \rangle \langle cd \rangle \langle bf \rangle + 
    \langle ae \rangle \langle cf \rangle \langle bd \rangle + 
    \langle af \rangle \langle cd \rangle \langle be \rangle + 
    \langle af \rangle \langle ce \rangle \langle bd \rangle \\ 
    &= \mathcal{I}\bigg[ 
    \big[ W_R(p) + W_R(|\mathbf{q}-\mathbf{p}|) \big] 
    \big[ W_R(p') + W_R(|\mathbf{q}'-\mathbf{p}'|) \big] 
    \left\langle \delta_{-\mathbf{q}} \delta_{\mathbf{p}'} \right\rangle
    \left\langle \delta_{\mathbf{q}-\mathbf{p}}\delta_{-\mathbf{q}'} \right \rangle
    \left\langle \delta_{\mathbf{p}}\delta_{\mathbf{q}'-\mathbf{p}'} \right\rangle \bigg]
\end{align}
showing that we only need to evaluate one contraction. Here, we have used $\tilde N_{222}$ to summarize the terms, as these are the non-diagonal contribution from the $(2,2,2)$ contractions.
Expanding this term, we find
\begin{align}
    & \tilde N_{222}  \nonumber\\
    &= \mathcal{I}\bigg[ \big[ W_R(p) + W_R(|\mathbf{q}-\mathbf{p}|) \big] 
    \big[ W_R(p') + W_R(|\mathbf{q}'-\mathbf{p}'|) \big] 
    \left\langle \delta_{-\mathbf{q}} \delta_{-\mathbf{p}'} \right\rangle
    \left\langle \delta_{\mathbf{q}-\mathbf{p}}\delta_{-\mathbf{q}'} \right \rangle
    \left\langle \delta_{\mathbf{p}}\delta_{\mathbf{q}'-\mathbf{p}'} \right\rangle \bigg] \\
    &= (2\pi)^9 V^{-2} \int_{V_k}\frac{d^3q}{V_k}\int_{V_k}\frac{d^3q'}{V_k} 
    \int_{\mathbf{p}}\int_{\mathbf{p}'} 
    \big[ W_R(p) + W_R(|\mathbf{q}-\mathbf{p}|) \big] 
    \big[ W_R(p') + W_R(|\mathbf{q}'-\mathbf{p}'|) \big]
    \nonumber \\
    &\qquad\qquad\qquad\qquad\qquad\qquad  \delta_D(\mathbf{q}-\mathbf{p}')
    \delta_D(\mathbf{q}-\mathbf{p}-\mathbf{q}') 
    \delta_D(\mathbf{p}+\mathbf{q}'-\mathbf{p}')
    P(-\mathbf{q})P(\mathbf{p})P(\mathbf{q}-\mathbf{p}) \\ 
    &= V^{-1} \int_{V_k}\frac{d^3q}{V_k}\int_{V_k}\frac{d^3q'}{V_k} 
    \big[ W_R(q) + W_R(|\mathbf{q}-\mathbf{q}'|) \big] 
    \big[ W_R(q') + W_R(|\mathbf{q}-\mathbf{q}'|) \big] \nonumber \\
    &\qquad\qquad\qquad\qquad\qquad\qquad\qquad\qquad\qquad\qquad\qquad\qquad\qquad\qquad\qquad P(\mathbf{q})P(\mathbf{q}')P(\mathbf{q}-\mathbf{q}')
\end{align}
making the $(2,2,2)$ covariance
\begin{align}
    \mathrm{Cov}\left(\widehat{P}_k, \widehat{M}_{k'}\right)_{222} &= (2\pi)^3 \delta^K_{k,k'} V^{-1} V_k^{-1} D_{222} + V^{-1}N_{222} \\
    D_{222}(\mathbf{k}) &= \int_\mathbf{p} W_R(p)\big[ W_R(p) + W_R(|\mathbf{k}-\mathbf{p}|) \big] P(\mathbf{k})P(\mathbf{p})P(\mathbf{k}-\mathbf{p}) \\
    N_{222}(\mathbf{k},\mathbf{k}') &= V \tilde N_{222} \\
    &= \left\langle
    \big[W_R(|\mathbf{k}-\mathbf{k}'|)+ W_R(k)\big]\big[W_R(|\mathbf{k}-\mathbf{k}'|)+ W_R(k')\big]  P(\mathbf{k})P(\mathbf{k}')P(\mathbf{k}-\mathbf{k}') \right \rangle_\varphi
\end{align}

Now let us consider the $(3,3)$ contractions. There are 10 possible terms
\begin{align}
\langle abcdef \rangle_{33} = 
\langle abc \rangle \langle def \rangle &+ \langle abd \rangle \langle cef \rangle + \langle abe \rangle \langle cdf \rangle + \langle abf \rangle \langle cde \rangle \, + \langle acd \rangle \langle bef \rangle + \nonumber \\
\langle ace \rangle \langle bdf \rangle &+ \langle acf \rangle \langle bde \rangle \, + \langle ade \rangle \langle bcf \rangle + \langle adf \rangle \langle bce \rangle \, + \langle aef \rangle \langle bcd \rangle
\end{align}
Immediately we notice that the first term involving $\expval{abc}$ cancels with the product of the spectra.
Let us attempt to group the remaining terms together by invoking the same transformation of the internal momenta $\mathbf{p}$ and $\mathbf{p}'$ as we did above. Specifically, we look for terms that are identical when performing $b\leftrightarrow c$ or $e\leftrightarrow f$, or both. 
First we recognize the following pair of terms
\begin{align}
    \langle abd \rangle \langle cef \rangle 
    &= \mathcal{I}\big[ W_R(p)W_R(p') 
    \left\langle \delta_{-\mathbf{q}} \delta_{\mathbf{p}} \delta_{-\mathbf{q}'} \right\rangle
    \left\langle \delta_{\mathbf{q}-\mathbf{p}} \delta_{\mathbf{p}'}\delta_{\mathbf{q}'-\mathbf{p}'} \right\rangle 
    \big] \\
    \langle acd  \rangle \langle bef\rangle 
    &= \mathcal{I}\big[ W_R(p)W_R(p') 
    \left\langle \delta_{-\mathbf{q}} \delta_{\mathbf{q}-\mathbf{p}} \delta_{-\mathbf{q}'} \right\rangle
    \left\langle \delta_{\mathbf{p}} \delta_{\mathbf{p}'}\delta_{\mathbf{q}'-\mathbf{p}'} \right\rangle 
    \big] \\
    &= \mathcal{I}\big[ W_R(|\mathbf{q}-\mathbf{p}|)W_R(p') 
    \left\langle \delta_{-\mathbf{q}} \delta_{\mathbf{p}} \delta_{-\mathbf{q}'} \right\rangle
    \left\langle \delta_{\mathbf{q}-\mathbf{p}} \delta_{\mathbf{p}'}\delta_{\mathbf{q}'-\mathbf{p}'} \right\rangle 
    \big] \\
    \langle abd \rangle \langle cef \rangle + \langle acd  \rangle \langle bef\rangle
    &= \mathcal{I}\bigg[ \big[ W_R(|\mathbf{q}-\mathbf{p}|) + W_R(p) \big]W_R(p')
    \left\langle \delta_{-\mathbf{q}} \delta_{\mathbf{p}} \delta_{-\mathbf{q}'} \right\rangle
    \left\langle \delta_{\mathbf{q}-\mathbf{p}} \delta_{\mathbf{p}'}\delta_{\mathbf{q}'-\mathbf{p}'} \right\rangle 
    \bigg]
    \label{eqn:CovM33pair1}
\end{align}
This is nearly identical to the pair
\begin{align}
    \langle ade \rangle \langle bcf \rangle 
    &= \mathcal{I}\big[ W_R(p)W_R(p') 
    \left\langle \delta_{-\mathbf{q}} \delta_{-\mathbf{q}'} \delta_{\mathbf{p}'} \right\rangle
    \left\langle \delta_{\mathbf{p}} \delta_{\mathbf{q}-\mathbf{p}} \delta_{\mathbf{q}'-\mathbf{p}'} \right\rangle 
    \big] \\
    \langle adf  \rangle \langle bce\rangle 
    &= \mathcal{I}\big[ W_R(p)W_R(p') 
    \left\langle \delta_{-\mathbf{q}} \delta_{-\mathbf{q}'} \delta_{\mathbf{q}'-\mathbf{p}'} \right\rangle
    \left\langle \delta_{\mathbf{p}} \delta_{\mathbf{q}-\mathbf{p}} \delta_{\mathbf{p}'} \right\rangle 
    \big] \\
    &= \mathcal{I}\big[ W_R(p)W_R(|\mathbf{q}'-\mathbf{p}'|) 
    \left\langle \delta_{-\mathbf{q}} \delta_{-\mathbf{q}'} \delta_{\mathbf{p}'} \right\rangle
    \left\langle \delta_{\mathbf{p}} \delta_{\mathbf{q}-\mathbf{p}} \delta_{\mathbf{q}'-\mathbf{p}'} \right\rangle 
    \big] \\
    \langle ade \rangle \langle bcf \rangle + \langle adf  \rangle \langle bce \rangle
    &= \mathcal{I}\bigg[\big[ W_R(|\mathbf{q}'-\mathbf{p}'|) + W_R(p')\big]W_R(p)
    \left\langle \delta_{-\mathbf{q}} \delta_{-\mathbf{q}'} \delta_{\mathbf{p}'} \right\rangle
    \left\langle \delta_{\mathbf{p}} \delta_{\mathbf{q}-\mathbf{p}} \delta_{\mathbf{q}'-\mathbf{p}'} \right\rangle 
    \bigg]
    \label{eqn:CovM33pair2}
\end{align}
other than the substitution $\mathbf{q}\leftrightarrow\mathbf{q}'$. 
Then, the quartet of terms
\begin{align}
    \langle abe \rangle \langle cdf \rangle
    &= \mathcal{I}\big[ W_R(p)W_R(p') 
    \left\langle \delta_{-\mathbf{q}} \delta_{\mathbf{p}} \delta_{\mathbf{p}'} \right\rangle
    \left\langle \delta_{\mathbf{q}-\mathbf{p}} \delta_{-\mathbf{q}'} \delta_{\mathbf{q}'-\mathbf{p}'} \right\rangle 
    \big] \\
    \langle abf \rangle \langle cde \rangle
    &= \mathcal{I}\big[ W_R(p)W_R(p') 
    \left\langle \delta_{-\mathbf{q}} \delta_{\mathbf{p}}  \delta_{\mathbf{q}'-\mathbf{p}'} \right\rangle
    \left\langle \delta_{\mathbf{q}-\mathbf{p}} \delta_{-\mathbf{q}'} \delta_{\mathbf{p}'} \right\rangle 
    \big] \\
    &= \mathcal{I}\big[ W_R(p)W_R(|\mathbf{q}'-\mathbf{p}'|) 
    \left\langle \delta_{-\mathbf{q}} \delta_{\mathbf{p}} \delta_{\mathbf{p}'} \right\rangle
    \left\langle \delta_{\mathbf{q}-\mathbf{p}} \delta_{-\mathbf{q}'} \delta_{\mathbf{q}'-\mathbf{p}'} \right\rangle 
    \big] \\
    \langle ace \rangle \langle bdf \rangle
    &= \mathcal{I}\big[ W_R(p)W_R(p') 
    \left\langle \delta_{-\mathbf{q}} \delta_{\mathbf{q}-\mathbf{p}} \delta_{\mathbf{p}'} \right\rangle
    \left\langle \delta_{\mathbf{p}} \delta_{-\mathbf{q}'} \delta_{\mathbf{q}'-\mathbf{p}'} \right\rangle 
    \big] \\
    &= \mathcal{I}\big[ W_R(|\mathbf{q}-\mathbf{p}|)W_R(p') 
    \left\langle \delta_{-\mathbf{q}} \delta_{\mathbf{p}} \delta_{\mathbf{p}'} \right\rangle
    \left\langle \delta_{\mathbf{q}-\mathbf{p}} \delta_{-\mathbf{q}'} \delta_{\mathbf{q}'-\mathbf{p}'} \right\rangle 
    \big] \\
    \langle acf \rangle \langle bde \rangle
    &= \mathcal{I}\big[ W_R(p)W_R(p') 
    \left\langle \delta_{-\mathbf{q}} \delta_{\mathbf{q}-\mathbf{p}} \delta_{\mathbf{q}'-\mathbf{p}'} \right\rangle
    \left\langle \delta_{\mathbf{p}} \delta_{-\mathbf{q}'} \delta_{\mathbf{p}'} \right\rangle 
    \big] \\
    &= \mathcal{I}\big[ W_R(|\mathbf{q}-\mathbf{p}|)W_R(|\mathbf{q}'-\mathbf{p}'|) 
    \left\langle \delta_{-\mathbf{q}} \delta_{\mathbf{p}} \delta_{\mathbf{p}'} \right\rangle
    \left\langle \delta_{\mathbf{q}-\mathbf{p}} \delta_{-\mathbf{q}'} \delta_{\mathbf{q}'-\mathbf{p}'} \right\rangle 
    \big] 
\end{align}
\begin{align}
    \langle abe \rangle & \langle cdf \rangle + \langle abf \rangle \langle cde \rangle + \langle ace \rangle \langle bdf \rangle + \langle acf \rangle \langle bde \rangle \nonumber \\
    &= \mathcal{I}\bigg[ \big[ W_R(p)+W_R(|\mathbf{q}-\mathbf{p}|)\big] 
    \big[  W_R(p') + W_R(|\mathbf{q}'-\mathbf{p}'|) \big]
    \left\langle \delta_{-\mathbf{q}} \delta_{\mathbf{p}} \delta_{\mathbf{p}'} \right\rangle
    \left\langle \delta_{\mathbf{q}-\mathbf{p}} \delta_{-\mathbf{q}'} \delta_{\mathbf{q}'-\mathbf{p}'} \right\rangle 
    \bigg] 
    \label{eqn:CovM33quartet}
\end{align}
Finally, the last term $\expval{aef}\expval{bcd}$ has a unique contraction. 

Let us start our evaluation with the last term $\expval{aef}\expval{bcd}$
\begin{align}
    \langle aef \rangle \langle bcd \rangle
    &= \mathcal{I}\big[ W_R(p)W_R(p') 
    \left\langle \delta_{-\mathbf{q}} \delta_{\mathbf{p}'} \delta_{\mathbf{q}'-\mathbf{p}'} \right\rangle
    \left\langle \delta_{\mathbf{p}} \delta_{\mathbf{q}-\mathbf{p}} \delta_{-\mathbf{q}'} \right\rangle 
    \big] \\
    &= (2\pi)^6 V^{-2} \int_{V_k}\frac{d^3q}{V_k}\int_{V_{k'}}\frac{d^3q'}{V_{k'}} 
    \int_{\mathbf{p}}\int_{\mathbf{p}'} W_R(p) W_R(p') \nonumber \\
    &\qquad\qquad \delta_D^2(-\mathbf{q}+\mathbf{q}') 
    B(-\mathbf{q},\mathbf{p}',\mathbf{q}'-\mathbf{p}')
    B(\mathbf{p},\mathbf{q}-\mathbf{p},-\mathbf{q}') \\
    &= (2\pi)^3 V^{-1} V_k^{-1} \delta^K_{k,k'} M(\mathbf{k}) M(\mathbf{k}')
\end{align}
which, we find, is the diagonal contribution. Next, the pairs of terms in Eqn.~\ref{eqn:CovM33pair1} and \ref{eqn:CovM33pair2}. As they are equivalent under $\mathbf{q}\leftrightarrow \mathbf{q}'$ ($\mathbf{k}\leftrightarrow \mathbf{k}'$), we only need to solve for the first pair
\begin{align}
    & \mathcal{I}\bigg[ \big[ W_R(|\mathbf{q}-\mathbf{p}|) + W_R(p) \big]W_R(p')
    \left\langle \delta_{-\mathbf{q}} \delta_{\mathbf{p}} \delta_{-\mathbf{q}'} \right\rangle
    \left\langle \delta_{\mathbf{q}-\mathbf{p}} \delta_{\mathbf{p}'}\delta_{\mathbf{q}'-\mathbf{p}'} \right\rangle 
    \bigg] \\
    &= (2\pi)^6 V^{-2} \int_{V_k}\frac{d^3q}{V_k}\int_{V_{k'}}\frac{d^3q'}{V_{k'}} 
    \int_{\mathbf{p}}\int_{\mathbf{p}'} \big[ W_R(|\mathbf{q}-\mathbf{p}|) + W_R(p) \big]W_R(p') \nonumber \\
    &\qquad\qquad \delta_D^2(\mathbf{q}+\mathbf{q}'-\mathbf{p}) B(-\mathbf{q},\mathbf{p},-\mathbf{q}') B(\mathbf{q}-\mathbf{p},\mathbf{p}',\mathbf{q}'-\mathbf{p}') \\
    &= V^{-1} \int_{V_k}\frac{d^3q}{V_k}\int_{V_{k'}}\frac{d^3q'}{V_{k'}} \int_{\mathbf{p}'} \big[ W_R(q') + W_R(|\mathbf{q}+\mathbf{q'}|) \big]W_R(p')  \nonumber \\
    &\qquad\qquad B(-\mathbf{q},\mathbf{q}+\mathbf{q}',-\mathbf{q}') B(-\mathbf{q}',\mathbf{p}',\mathbf{q}'-\mathbf{p}') \\
    &= V^{-1} \left\langle \big[ W_R(k') + W_R(|\mathbf{k}+\mathbf{k'}|) \big] B(-\mathbf{k},\mathbf{k}+\mathbf{k}',-\mathbf{k}') M(\mathbf{k}') \right\rangle_\varphi
\end{align}
Finally the quartet of terms in Eqn.~\ref{eqn:CovM33quartet}
\begin{align}
    & \mathcal{I}\bigg[ \big[ W_R(p)+W_R(|\mathbf{q}-\mathbf{p}|)\big] 
    \big[  W_R(p') + W_R(|\mathbf{q}'-\mathbf{p}'|) \big]
    \left\langle \delta_{-\mathbf{q}} \delta_{\mathbf{p}} \delta_{\mathbf{p}'} \right\rangle
    \left\langle \delta_{\mathbf{q}-\mathbf{p}} \delta_{-\mathbf{q}'} \delta_{\mathbf{q}'-\mathbf{p}'} \right\rangle 
    \bigg] \\
    &= (2\pi)^6 V^{-2} \int_{V_k}\frac{d^3q}{V_k}\int_{V_{k'}}\frac{d^3q'}{V_{k'}} 
    \int_{\mathbf{p}}\int_{\mathbf{p}'} \big[ W_R(p)+W_R(|\mathbf{q}-\mathbf{p}|)\big] 
    \big[  W_R(p') + W_R(|\mathbf{q}'-\mathbf{p}'|) \big] \nonumber \\
    &\qquad\qquad \delta_D^2(-\mathbf{q}+\mathbf{p}+\mathbf{p}') 
    B(-\mathbf{q},\mathbf{p},\mathbf{p}') B(\mathbf{q}-\mathbf{p},-\mathbf{q}',\mathbf{q}'-\mathbf{p}') \\
    &= V^{-1} \bigg\langle \int_\mathbf{p} \big[ W_R(p)+W_R(|\mathbf{k}-\mathbf{p}|)\big] 
    \big[  W_R(p) + W_R(|\mathbf{k}'-\mathbf{p}|) \big] \nonumber \\
    &\qquad\qquad  B(-\mathbf{k},\mathbf{p},\mathbf{k}-\mathbf{p}) B(\mathbf{p},-\mathbf{k}',\mathbf{k}'-\mathbf{p})
    \bigg\rangle_\varphi
\end{align}
where, in the last line we apply the transformation $\mathbf{p}\to \mathbf{k}-\mathbf{p}$ along with the thin-shell approximation.

Let us now consider the $(4,2)$ contributions. There are 15 possible terms, that can be reduced to 9 terms using the same method we had used above for $(2,2,2)$
\begin{align}
\langle abcdef \rangle_{42} = 
\langle ad\rangle\langle bcef\rangle &+ \langle ae\rangle\langle bcdf\rangle + \langle af\rangle\langle bcde\rangle +
\langle bd\rangle\langle acef\rangle + \langle be\rangle\langle acdf\rangle + \nonumber \\
\langle bf\rangle\langle acde\rangle &+ \langle cd\rangle\langle abef\rangle +
\langle ce\rangle\langle abdf\rangle + \langle cf\rangle\langle abde\rangle
\end{align}
Using the transformation $\mathbf{p}\to\mathbf{q}-\mathbf{p}$ we can group terms together
\begin{align}
    \langle ae\rangle\langle bcdf\rangle + \langle af\rangle\langle bcde\rangle &= \mathcal{I}\bigg[W_R(p)\left[W_R(p')+W_R(|\mathbf{q}'-\mathbf{p}'|) \right]
    \langle \delta_{-\mathbf{q}}\delta_{\mathbf{p}'} \rangle
    \langle \delta_\mathbf{p} \delta_{\mathbf{q}-\mathbf{p}} \delta_{-\mathbf{q}'} \delta_{\mathbf{q}'-\mathbf{p}'} \rangle
    \bigg]
\end{align}
Similarly, $\langle bd\rangle\langle acef\rangle +\langle cd\rangle\langle abef\rangle$ yield the equivalent expression with $\mathbf{k}\leftrightarrow\mathbf{k}'$. The terms
\begin{align}
    &\langle be\rangle\langle acdf\rangle + \langle bf\rangle\langle acde\rangle +
    \langle ce\rangle\langle abdf\rangle + \langle cf\rangle\langle abde\rangle \nonumber \\
    & \quad = \mathcal{I}\bigg[\left[W_R(p)+W_R(|\mathbf{q}-\mathbf{p}|)\right] \left[W_R(p')+W_R(|\mathbf{q}'-\mathbf{p}'|)\right] 
    \langle \delta_{\mathbf{p}}\delta_{\mathbf{p}'} \rangle
    \langle \delta_{-\mathbf{q}} \delta_{\mathbf{q}-\mathbf{p}} \delta_{-\mathbf{q}'} \delta_{\mathbf{q}'-\mathbf{p}'} \rangle
      \bigg]
\end{align}
also arise from equivalent contractions. With the unique term $\langle ad\rangle\langle bcef\rangle$, we have three unique contractions to evaluate.
Each of these terms become
\begin{align}
    \langle ad\rangle\langle bcef\rangle = (2\pi)^3 V^{-1} V_k^{-1} \delta^K_{k,k'}  \int_\mathbf{p} \int_{\mathbf{p}'} W_R(p)W_R(p')P(\mathbf{k})
    T(\mathbf{p},\mathbf{k}-\mathbf{p},\mathbf{p}',-\mathbf{k}-\mathbf{p}')
\end{align}
\begin{align}
    & \langle ae\rangle\langle bcdf\rangle + \langle af\rangle\langle bcde\rangle \nonumber \\
    &= V^{-1} \left\langle \int_\mathbf{p} W_R(p) [W_R(k)+W_R(|\mathbf{k}-\mathbf{k}'|)] P(\mathbf{k}) T(\mathbf{p},\mathbf{k}-\mathbf{p},-\mathbf{k}',\mathbf{k}'-\mathbf{k}) 
    \right\rangle_\varphi
\end{align}
\begin{align}
    &\langle be\rangle\langle acdf\rangle + \langle bf\rangle\langle acde\rangle +
    \langle ce\rangle\langle abdf\rangle + \langle cf\rangle\langle abde\rangle \nonumber \\
    &= V^{-1} \left\langle \int_\mathbf{p} [W_R(k)+W_R(|\mathbf{k}-\mathbf{k}'|)][W_R(k')+W_R(|\mathbf{k}-\mathbf{k}'|)] P_\mathbf{p} T(-\mathbf{k},\mathbf{k}-\mathbf{p},-\mathbf{k}',\mathbf{k}'+\mathbf{p}) 
    \right\rangle_\varphi
\end{align}

This leaves us to evaluate the final term in the covariance, contracting all 6 fields together. This yields
\begin{align}
    \mathrm{Cov}\left(\widehat{M}_k,\widehat{M}_k'\right)_{6} = \langle abcdef\rangle &= V^{-1} \left\langle \int_\mathbf{p} \int_{\mathbf{p}'} 
    W_R(p)W_R(p')P_6(-\mathbf{k},\mathbf{p},\mathbf{k}-\mathbf{p},-\mathbf{k}',\mathbf{p}',\mathbf{k}'-\mathbf{p}')
    \right\rangle_\varphi
\end{align}

The final expression for the covariance becomes
\begin{align}
    \mathrm{Cov}\left(\widehat{M}_k,\widehat{M}_k'\right) &= (2\pi)^3 \delta^K_{k,k'} V^{-1} V_k^{-1} \left[ D_{22}+D_{33}+D_{42} \right] + 
    V^{-1}[N_{222}+N_{33}+N_{42}+N_{6}] 
\end{align}
with diagonal terms
\begin{align}
    D_{222}(\mathbf{k}) &= \int_\mathbf{p} W_R(p)\big[ W_R(p) + W_R(|\mathbf{k}-\mathbf{p}|) \big] P(\mathbf{k})P(\mathbf{p})P(\mathbf{k}-\mathbf{p}) 
    \label{eqn:D222} \\    
    D_{33}(\mathbf{k}) &= M(\mathbf{k})^2 \\
    D_{42}(\mathbf{k}) &= \int_\mathbf{p} \int_{\mathbf{p}'} W_R(p)W_R(p')P(\mathbf{k})
    T(\mathbf{p},\mathbf{k}-\mathbf{p},\mathbf{p}',-\mathbf{k}-\mathbf{p}')
\end{align}
and non-diagonal terms
\begin{align}
    N_{222}(\mathbf{k},\mathbf{k}') 
    &= \left\langle
    \big[W_R(|\mathbf{k}-\mathbf{k}'|)+ W_R(k)\big]\big[W_R(|\mathbf{k}-\mathbf{k}'|)+ W_R(k')\big]  P(\mathbf{k})P(\mathbf{k}')P(\mathbf{k}-\mathbf{k}') \right \rangle_\varphi \\
    N_{33}(\mathbf{k},\mathbf{k}') &= \left\langle \big[ W_R(k') + W_R(|\mathbf{k}+\mathbf{k'}|) \big] B(-\mathbf{k},\mathbf{k}+\mathbf{k}',-\mathbf{k}') M(\mathbf{k}') \right\rangle_\varphi \nonumber \\ 
    &\quad + \bigg\langle \int_\mathbf{p} \big[ W_R(p)+W_R(|\mathbf{k}-\mathbf{p}|)\big] 
    \big[  W_R(p) + W_R(|\mathbf{k}'-\mathbf{p}|) \big] \nonumber \\
    &\qquad\qquad  B(-\mathbf{k},\mathbf{p},\mathbf{k}-\mathbf{p}) B(\mathbf{p},-\mathbf{k}',\mathbf{k}'-\mathbf{p})
    \bigg\rangle_\varphi \\
    N_{42}(\mathbf{k},\mathbf{k}') &=  \left\langle \int_\mathbf{p} W_R(p) [W_R(k)+W_R(|\mathbf{k}-\mathbf{k}'|)] P(\mathbf{k}) T(\mathbf{p},\mathbf{k}-\mathbf{p},-\mathbf{k}',\mathbf{k}'-\mathbf{k}) 
    \right\rangle_\varphi \nonumber \\
    &\quad + \left\langle \int_\mathbf{p} [W_R(k)+W_R(|\mathbf{k}-\mathbf{k}'|)][W_R(k')+W_R(|\mathbf{k}-\mathbf{k}'|)] P_\mathbf{p} T(-\mathbf{k},\mathbf{k}-\mathbf{p},-\mathbf{k}',\mathbf{k}'+\mathbf{p}) 
    \right\rangle_\varphi \\
    N_{6}(\mathbf{k},\mathbf{k}') &= \left\langle \int_\mathbf{p} \int_{\mathbf{p}'} 
    W_R(p)W_R(p')P_6(-\mathbf{k},\mathbf{p},\mathbf{k}-\mathbf{p},-\mathbf{k}',\mathbf{p}',\mathbf{k}'-\mathbf{p}')
    \right\rangle_\varphi
\end{align}

Before moving on to the $P$-$M$ covariance, let us compare this with the result under the Gaussian approximation. The Gaussian (disconnected) covariance can be computed by assuming that each field involved ($\delta$ and $\tilde\delta_M$) is Gaussian. Invoking Wick's theorem \cite{Wick50}, one finds
\begin{align}
    \mathrm{Cov}_G\left(\widehat{P}_k, \widehat{M}_{k'}\right) 
    &= V^{-2} \int_{V_k} \frac{d^3q}{V_k} \int_{V_k} \frac{d^3q'}{V_k} 
    \left(\left\langle \tilde\delta_{M,-q}\delta_q \tilde\delta_{M,-q'} \delta_{q'} \right\rangle
    - \left\langle \tilde\delta_{M,-q}\delta_q \right\rangle
    \left\langle \tilde\delta_{M,-q'} \delta_{q'} \right\rangle \right)\\ 
    &= V^{-2} \int_{V_k} \frac{d^3q}{V_k} \int_{V_k} \frac{d^3q'}{V_k} 
    \left\langle \tilde\delta_{M,-q}\delta_{q'} \right \rangle
    \left\langle\delta_k \tilde\delta_{M,-q'}  \right\rangle 
    + \left\langle \tilde\delta_{M,-q}\tilde\delta_{M,-q'} \right \rangle
    \left\langle\delta_q \delta_{q'}  \right\rangle \\
    &= (2\pi)^3 \delta^K_{k,k'} V^{-1} V_{k}^{-1} \left [M^2(\mathbf{k})
    + P(\mathbf{k})V^{-1} \left\langle \tilde\delta_{M,k}\tilde\delta_{M,-k} \right \rangle
      \right ]
\end{align}
The only term above that cannot be expressed as $P$ or $M$ is $\left\langle \tilde\delta_{M,k}\tilde\delta_{M,-k} \right \rangle$. 
Expanding this, we find
\begin{align}
    \left\langle \tilde\delta_{M,-k}\tilde\delta_{M,-k} \right \rangle 
    &= \int_{V_{k}} \frac{d^3q}{V_k}\, \left\langle\tilde\delta_{M,\mathbf{q}}\tilde\delta_{M,-\mathbf{q}}\right \rangle  \\ 
    &= \int_{V_{k}} \frac{d^3q}{V_k} \int_\mathbf{p} \int_{\mathbf{p}'} W_R(p) W_R(p') \left\langle \delta_\mathbf{p} \delta_{\mathbf{q}-\mathbf{p}}
    \delta_{\mathbf{p}'} \delta_{-\mathbf{q}-\mathbf{p}'}
    \right \rangle 
\end{align}
The four-point correlator contributes both $(2,2)$ disconnected contractions and a connected four-point contraction. One of the disconnected terms, involving $\langle \delta_\mathbf{p} \delta_{\mathbf{q}-\mathbf{p}} \rangle$ vanishes. The other disconnected terms contribute
\begin{align}
    \left\langle \tilde\delta_{M,-k} \tilde\delta_{M,-k} \right\rangle_{22}
    &= V \int_{V_{k}} \frac{d^3q}{V_k} \int_\mathbf{p} W_R(p) \left [W_R(p) + W_R(|\mathbf{q}-\mathbf{p}|) \right ] P(\mathbf{q})P(\mathbf{q}-\mathbf{p})  \\    
    &= V \int_\mathbf{p} W_R(p) \left [W_R(p) + W_R(|\mathbf{k}-\mathbf{p}|) \right ] P(\mathbf{k})P(\mathbf{k}-\mathbf{p}) 
\end{align}
while the connected piece yields
\begin{align}
    \left\langle \tilde\delta_{M,-k} \tilde\delta_{M,-k} \right\rangle_{4}
    &= \int_{V_{k}} \frac{d^3q}{V_k} \int_\mathbf{p} \int_{\mathbf{p}'} W_R(p) W_R(p') T(\mathbf{p},\mathbf{q}-\mathbf{p}, \mathbf{p}',-\mathbf{q}-\mathbf{p}')
\end{align}
Thus the Gaussian covariance becomes 
\begin{align}
    \mathrm{Cov}_G\left(\widehat{P}_k, \widehat{M}_{k'}\right) &= (2\pi)^3 \delta^K_{k,k'} V^{-1} V_k^{-1} [D_{222}(k) + D_{42}(k) + D_{33}(k)]
\end{align}
indicating that the Gaussian approximation correctly captures the full diagonal contribution, which is dominant at large scales. 

\subsection{Covariance of $P-M$}
\label{app:covariance_sub}

We now calculate the cross-covariance between $P$ and $M$ (\S\ref{sec:crosscov}). The expression for the cross-covariance is
\begin{align}
    \mathrm{Cov}\left(\widehat{P}_k, \widehat{M}_{k'}\right) &= 
    V^{-2} \int_{V_k} \frac{d^3q}{V_k}  \int_{V_k} \frac{d^3q'}{V_k} \int_{\mathbf{p}'} W_R(p') \\
    &\qquad (\left\langle \delta_{-\mathbf{q}} \delta_{\mathbf{q}} \delta_{-\mathbf{q}'} \delta_{\mathbf{p}'} \delta_{\mathbf{q}'-\mathbf{p}'}\right\rangle 
    - \left\langle\delta_{-\mathbf{q}} \delta_{\mathbf{q}}\right\rangle 
    \left\langle\delta_{-\mathbf{q}'} \delta_{\mathbf{p}'} \delta_{\mathbf{q}'-\mathbf{p}'} \right\rangle) \\
    &= \mathcal{I}[W_R(p') \left\langle \delta_{-\mathbf{q}} \delta_{\mathbf{q}} \delta_{-\mathbf{q}'} \delta_{\mathbf{p}'} \delta_{\mathbf{q}'-\mathbf{p}'}\right\rangle ] - P_k M_{k'}
\end{align}
where $\mathcal{I}$ is now 
\begin{equation}
    \mathcal{I}[f] = V^{-2}\int_{V_k} \frac{d^3q\,d^3q'}{V_k^2} \int_{\mathbf{p}'} f(\mathbf{q},\mathbf{q}',\mathbf{p}')
\end{equation}
We again start by considering the possible permutations of the five-point correlator. As $\langle\delta\rangle=0$, we are restricted to considering contractions of the form $(3,2)$ or $(5)$. There are 10 possible contractions of the $(32)$ form
\begin{align}
    \langle abcde\rangle_{32} =&\quad \langle ab\rangle\langle cde\rangle + \langle ac\rangle\langle bde\rangle + \langle ad\rangle\langle bce\rangle + \langle ae\rangle\langle bcd\rangle + \langle bc\rangle\langle ade\rangle \nonumber \\
    + &\quad \langle bd\rangle\langle ace\rangle + \langle be\rangle\langle acd\rangle + \langle cd\rangle\langle abe\rangle + \langle ce\rangle\langle abd\rangle + \langle de\rangle\langle abc\rangle
\end{align}
\begin{equation}
    \mathrm{with}\qquad\qquad
    a = \delta_{-\mathbf{q}},\quad b=\delta_{\mathbf{q}},\quad c=\delta_{-\mathbf{q}'},\quad
    d = \delta_{\mathbf{p}'},\quad e=\delta_{\mathbf{q}'-\mathbf{p}'}
\end{equation}
Immediately we notice that the first term involving $\langle ab\rangle$ cancels with the product of the spectra. Of the remaining terms, we again take advantage of the property that the momenta of $c=\delta_{-\mathbf{q}'}$, $d = \delta_{\mathbf{p}'}$, and $ e=\delta_{\mathbf{q}'-\mathbf{p}'}$ sum to zero, hence having an ensemble average over two of the fields results in the third field being evaluated at zero momentum. This eliminates the last three terms, leaving us with six. Then, notice that $a = \delta_{-\mathbf{q}}$ and $ b=\delta_{\mathbf{q}}$ merely differ by the sign of $\mathbf{q}$, which can be dropped by symmetry. 
Thus the final contribution will be identical to 
\begin{equation}
    2 \left[
    \langle ac \rangle \langle bde \rangle + \langle ad \rangle \langle bce \rangle + \langle ae \rangle \langle bcd \rangle \right] 
\end{equation}
Let us start the evaluation with 
\begin{align}
    \mathcal{I}[W_R(p')\langle ac \rangle\langle bde \rangle] 
    &= \mathcal{I}[W_R(p')\langle \delta_{-\mathbf{q}}\delta_{-\mathbf{q}'} \rangle \langle \delta_{\mathbf{q}}\delta_{\mathbf{p}'}\delta_{\mathbf{q}'-\mathbf{p}'} \rangle] \\
    &= \mathcal{I}[(2\pi)^6 W_R(p') \delta_D^2(\mathbf{q}+\mathbf{q}') B(\mathbf{p}',\mathbf{q}-\mathbf{p}',-\mathbf{q})P(-\mathbf{q})] \\
    &= V^{-1} V_k^{-1} \delta^K_{k,-k'} P(-\mathbf{k}) \int_{\mathbf{p}'} W_R(p') B(\mathbf{p}',\mathbf{k}-\mathbf{p}',-\mathbf{k}) \\
    &= V^{-1} V_k^{-1} \delta^K_{k,k'} P(\mathbf{k}) M(\mathbf{k})
\end{align}
where in the last line we drop the sign of $\mathbf{k}$ due to symmetry. We notice that this is the diagonal contribution to the covariance. 
Next, we have 
\begin{align}
    \mathcal{I}[W_R(p')\langle ad \rangle\langle bce \rangle] 
    &= \mathcal{I}[W_R(p')\langle \delta_{-\mathbf{q}}\delta_{\mathbf{p}'} \rangle \langle \delta_{\mathbf{q}}\delta_{-\mathbf{q}'}\delta_{\mathbf{q}'-\mathbf{p}'} \rangle] \\
    &= \mathcal{I}[(2\pi)^6 W_R(p') \delta_D^2(\mathbf{q}-\mathbf{p}') B(\mathbf{q},-\mathbf{q}',\mathbf{q}'-\mathbf{p}')P(-\mathbf{q})] \\
    &= V^{-1} V_k^{-2} \int_{V_k} d^3q  \int_{V_k} d^3q' W_R(q) P(\mathbf{q}) B(\mathbf{q},-\mathbf{q}',\mathbf{q}'-\mathbf{q}) \\
    &= V^{-1} W_R(k) P(\mathbf{k}) \left\langle B(\mathbf{k},-\mathbf{k}',\mathbf{k}'-\mathbf{k}) \right\rangle_\varphi
\end{align}
Finally, 
\begin{align}
    \mathcal{I}[W_R(p')\langle ae \rangle\langle bcd\rangle] 
    &= \mathcal{I}[W_R(p')\langle \delta_{-\mathbf{q}}\delta_{\mathbf{q}'-\mathbf{p}'} \rangle \langle \delta_{\mathbf{q}}\delta_{-\mathbf{q}'}\delta_{\mathbf{p}'} \rangle] \\
    &= \mathcal{I}[(2\pi)^6 W_R(p') \delta_D^2(\mathbf{q}-\mathbf{q}'+\mathbf{p}') B(\mathbf{q},-\mathbf{q}',\mathbf{p}')P(-\mathbf{q})] \\
    &= V^{-1} V_k^{-2} \int_{V_k} d^3q  \int_{V_k} d^3q' W_R(|\mathbf{q}'-\mathbf{q}|) P(\mathbf{q}) B(\mathbf{q},-\mathbf{q}',\mathbf{q}'-\mathbf{q}) \\
    &= V^{-1} P(\mathbf{k}) \left\langle W_R(|\mathbf{k}'-\mathbf{k}|) B(\mathbf{k},-\mathbf{k}',\mathbf{k}'-\mathbf{k}) \right\rangle_\varphi
\end{align}

Now we consider the only other contribution, contracting all five operators
\begin{align}
    \mathcal{I}[W_R(p')\langle abcde\rangle] &=  
    \mathcal{I}[W_R(p')\left\langle \delta_{-\mathbf{q}} \delta_{\mathbf{q}} \delta_{-\mathbf{q}'} \delta_{\mathbf{p}'} \delta_{\mathbf{q}'-\mathbf{p}'}\right\rangle]\\
    &= \mathcal{I}\left[(2\pi)^3 W_R(p') \delta_D(0)P_5\left(-\mathbf{q},\mathbf{q},-\mathbf{q}',\mathbf{p}',\mathbf{q}'-\mathbf{p}'\right)\right] \\
    &= V^{-1} V_k^{-2} \int_{V_k} d^3q  \int_{V_k} d^3q' \int_{\mathbf{p}'} W_R(p') P_5\left(-\mathbf{q},\mathbf{q},-\mathbf{q}',\mathbf{p}',\mathbf{q}'-\mathbf{p}'\right) \\
    &= V^{-1} \left\langle \int_{\mathbf{p}'} W_R(p') P_5\left(-\mathbf{k},\mathbf{k},-\mathbf{k}',\mathbf{p}',\mathbf{k}'-\mathbf{p}'\right) \right\rangle_\varphi
\end{align}

Thus the final contributions are 
\begin{align}
    \mathrm{Cov}\left(\widehat{P}_k, \widehat{M}_{k'}\right) &= (2\pi)^3 \delta^K_{k,k'} V^{-1} V_k^{-1} D_{23} + V^{-1}\left[N_{23} + N_5  \right] \\
    D_{23}(\mathbf{k}) &= 2 P(\mathbf{k}) M(\mathbf{k}) 
    \label{eqn:D23} \\
    N_{23}(\mathbf{k},\mathbf{k}') &=  2 P(\mathbf{k}) \left\langle (W_R(k) +  W_R(|\mathbf{k}'-\mathbf{k}|))B(\mathbf{k},-\mathbf{k}',\mathbf{k}'-\mathbf{k}) \right\rangle_\varphi\\
    N_5 &= 2 \left\langle \int_{\mathbf{p}'} W_R(p') P_5\left(-\mathbf{k},\mathbf{k},-\mathbf{k}',\mathbf{p}',\mathbf{k}'-\mathbf{p}'\right) \right\rangle_\varphi 
\end{align}
As with the auto-covariance, we observe that the disconnected term dominates at large-scales due to the $V_k^{-1}$ scaling, but the non-diagonal terms become important as we go to small scales. 

Now we compute the cross-covariance assuming that the fields involved are Gaussian. Using Wick's theorem \cite{Wick50,Harscouet24}, 
\begin{align}
    \mathrm{Cov}_G\left(\widehat{P}_k, \widehat{M}_{k'}\right) 
    &= V^{-2} \int_{V_k} \frac{d^3q}{V_k} \int_{V_k} \frac{d^3q'}{V_k} 
    \left(\left\langle \delta_{-q}\delta_q \tilde\delta_{M,-q'} \delta_{q'} \right\rangle
    - \left\langle \delta_{-q}\delta_q \right\rangle
    \left\langle \tilde\delta_{M,-q'} \delta_{q'} \right\rangle \right)\\ 
    &= V^{-2} \int_{V_k} \frac{d^3q}{V_k} \int_{V_k} \frac{d^3q'}{V_k} 
    \left\langle \delta_{-q}\delta_{q'} \right \rangle
    \left\langle\delta_q \tilde\delta_{M,-q'}  \right\rangle 
    + \left\langle \delta_{-q}\tilde\delta_{M,-q'} \right \rangle
    \left\langle\delta_q \delta_{q'}  \right\rangle\\ 
    &= (2\pi)^3\delta^K_{k,k'} V^{-1} V_k^{-1} 2 P(\mathbf{k}) M(\mathbf{k}') \\
    &= (2\pi)^3\delta^K_{k,k'} V^{-1} V_k^{-1} D_{23}
\end{align}
Thus we find again that the Gaussian approximation recovers the diagonal entries of the covariance matrix. At the large-scales that $M$ is applicable, these terms are expected to dominate.

One can extend these calculations to redshift space by utilizing the multipole expansion of power spectra when invoking the thin-shell approximation:  $P(\mathbf{q})=\sum_\ell P(q)\mathcal{L}_\ell(\mu) \approx \sum_\ell P(k)\mathcal{L}_\ell(\mu) = P(\mathbf{k})$. 

\bibliographystyle{JHEP}
\bibliography{main}

\end{document}